\documentclass[longauth]{aa}
\usepackage{txfonts}
\usepackage{graphicx}
\usepackage{float}
\usepackage{natbib}
\usepackage[english]{babel}
\usepackage[mediumspace,Gray,squaren]{SIunits}

\usepackage{color}

\title{HerMES: deep number counts at 250\,$\mu$m, 350\,$\mu$m and 500\,$\mu$m in the COSMOS and GOODS-N fields and the build-up of the cosmic infrared background}

\author{M.~B{\'e}thermin\inst{1,2}
\and E.~Le Floc'h\inst{1}
\and O.~Ilbert\inst{3}
\and A.~Conley\inst{4}
\and G.~Lagache\inst{2}
\and A.~Amblard\inst{5}
\and V.~Arumugam\inst{6}
\and H.~Aussel\inst{1}
\and S.~Berta\inst{7}
\and J.~Bock\inst{8,9}
\and A.~Boselli\inst{3}
\and V.~Buat\inst{3}
\and C.M.~Casey\inst{10}
\and N.~Castro-Rodr{\'\i}guez\inst{11,12}
\and A.~Cava\inst{13}
\and D.L.~Clements\inst{14}
\and A.~Cooray\inst{15,8}
\and C.D.~Dowell\inst{8,9}
\and S.~Eales\inst{16}
\and D.~Farrah\inst{17}
\and A.~Franceschini\inst{18}
\and J.~Glenn\inst{19,4}
\and M.~Griffin\inst{16}
\and E.~Hatziminaoglou\inst{20}
\and S.~Heinis\inst{3}
\and E.~Ibar\inst{21}
\and R.J.~Ivison\inst{21,6}
\and J.S.~Kartaltepe\inst{22,23}
\and L.~Levenson\inst{8,9}
\and G.~Magdis\inst{1}
\and L.~Marchetti\inst{18}
\and G.~Marsden\inst{24}
\and H.T.~Nguyen\inst{9,8}
\and B.~O'Halloran\inst{14}
\and S.J.~Oliver\inst{17}
\and A.~Omont\inst{25}
\and M.J.~Page\inst{26}
\and P.~Panuzzo\inst{1}
\and A.~Papageorgiou\inst{16}
\and C.P.~Pearson\inst{27,28}
\and I.~P{\'e}rez-Fournon\inst{11,12}
\and M.~Pohlen\inst{16}
\and D.~Rigopoulou\inst{27,29}
\and I.G.~Roseboom\inst{17,6}
\and M.~Rowan-Robinson\inst{14}
\and M.~Salvato\inst{30}
\and B.~Schulz\inst{8,31}
\and Douglas~Scott\inst{24}
\and N.~Seymour\inst{32,26}
\and D.L.~Shupe\inst{8,31}
\and A.J.~Smith\inst{17}
\and M.~Symeonidis\inst{26}
\and M.~Trichas\inst{33}
\and K.E.~Tugwell\inst{26}
\and M.~Vaccari\inst{18}
\and I.~Valtchanov\inst{34}
\and J.D.~Vieira\inst{8}
\and M.~Viero\inst{8}
\and L.~Wang\inst{17}
\and C.K.~Xu\inst{8,31}
\and M.~Zemcov\inst{8,9}}
\institute{Laboratoire AIM-Paris-Saclay, CEA/DSM/Irfu - CNRS - Universit\'e Paris Diderot, CE-Saclay, pt courrier 131, F-91191 Gif-sur-Yvette, France\\
\email{matthieu.bethermin@cea.fr}
\and Institut d'Astrophysique Spatiale (IAS), b\^atiment 121, Universit\'e Paris-Sud 11 and CNRS (UMR 8617), 91405 Orsay, France
\and Laboratoire d'Astrophysique de Marseille, OAMP, Universit\'e Aix-marseille, CNRS, 38 rue Fr\'ed\'eric Joliot-Curie, 13388 Marseille cedex 13, France
\and Center for Astrophysics and Space Astronomy 389-UCB, University of Colorado, Boulder, CO 80309, USA
\and NASA, Ames Research Center, Moffett Field, CA 94035, USA
\and Institute for Astronomy, University of Edinburgh, Royal Observatory, Blackford Hill, Edinburgh EH9 3HJ, UK
\and Max-Planck-Institut f\"ur Extraterrestrische Physik (MPE), Postfach 1312, 85741, Garching, Germany
\and California Institute of Technology, 1200 E. California Blvd., Pasadena, CA 91125, USA
\and Jet Propulsion Laboratory, 4800 Oak Grove Drive, Pasadena, CA 91109, USA
\and Institute for Astronomy, University of Hawaii, 2680 Woodlawn Drive, Honolulu, HI 96822, USA
\and Instituto de Astrof{\'\i}sica de Canarias (IAC), E-38200 La Laguna, Tenerife, Spain
\and Departamento de Astrof{\'\i}sica, Universidad de La Laguna (ULL), E-38205 La Laguna, Tenerife, Spain
\and Departamento de Astrof\'isica, Facultad de CC. F\'isicas, Universidad Complutense de Madrid, E-28040 Madrid, Spain
\and Astrophysics Group, Imperial College London, Blackett Laboratory, Prince Consort Road, London SW7 2AZ, UK
\and Dept. of Physics \& Astronomy, University of California, Irvine, CA 92697, USA
\and School of Physics and Astronomy, Cardiff University, Queens Buildings, The Parade, Cardiff CF24 3AA, UK
\and Astronomy Centre, Dept. of Physics \& Astronomy, University of Sussex, Brighton BN1 9QH, UK
\and Dipartimento di Astronomia, Universit\`{a} di Padova, vicolo Osservatorio, 3, 35122 Padova, Italy
\and Dept. of Astrophysical and Planetary Sciences, CASA 389-UCB, University of Colorado, Boulder, CO 80309, USA
\and ESO, Karl-Schwarzschild-Str. 2, 85748 Garching bei M\"unchen, Germany
\and UK Astronomy Technology Centre, Royal Observatory, Blackford Hill, Edinburgh EH9 3HJ, UK
\and Hubble Fellow
\and National Optical Astronomy Observatory, 950 North Cherry Avenue, Tucson, AZ 85719, USA
\and Department of Physics \& Astronomy, University of British Columbia, 6224 Agricultural Road, Vancouver, BC V6T~1Z1, Canada
\and Institut d'Astrophysique de Paris, UMR 7095, CNRS, UPMC Univ. Paris 06, 98bis boulevard Arago, F-75014 Paris, France
\and Mullard Space Science Laboratory, University College London, Holmbury St. Mary, Dorking, Surrey RH5 6NT, UK
\and RAL Space, Rutherford Appleton Laboratory, Chilton, Didcot, Oxfordshire OX11 0QX, UK
\and Institute for Space Imaging Science, University of Lethbridge, Lethbridge, Alberta, T1K 3M4, Canada
\and Department of Astrophysics, Denys Wilkinson Building, University of Oxford, Keble Road, Oxford OX1 3RH, UK
\and Max Planck Institut f\"ur Plasma Physik and Excellence Cluster, 85748 Garching, Germany
\and Infrared Processing and Analysis Center, MS 100-22, California Institute of Technology, JPL, Pasadena, CA 91125, USA
\and CSIRO Astronomy \& Space Science, PO Box 76, Epping, NSW 1710, Australia
\and Harvard-Smithsonian Center for Astrophysics, 60 Garden Street, Cambridge, MA 02138, USA
\and Herschel Science Centre, European Space Astronomy Centre, Villanueva de la Ca\~nada, 28691 Madrid, Spain}

\date{Received ??? / Accepted ???}

\abstract{}{
The Spectral and Photometric Imaging REceiver (SPIRE) onboard the \textit{Herschel} space telescope has provided confusion limited maps of deep fields at 250\,$\mu$m, 350\,$\mu$m, and 500\,$\micro$m, as part of the \textit{Herschel} Multi-tiered Extragalactic Survey (HerMES). Unfortunately, due to confusion, only a small fraction of the cosmic infrared background (CIB) can be resolved into individually-detected sources. Our goal is to produce deep galaxy number counts and redshift distributions below the confusion limit at SPIRE wavelengths ($\sim$20\,mJy), which we then use to place strong constraints on the origins of the cosmic infrared background and on models of galaxy evolution.
}{
We individually extracted the bright SPIRE sources ($>$20\,mJy) in the COSMOS field with a method using the positions, the flux densities, and the redshifts of the 24~$\mu$m sources as a prior, and derived the number counts and redshift distributions of the bright SPIRE sources. For fainter SPIRE sources ($<$20\,mJy), we reconstructed the number counts and the redshift distribution below the confusion limit using the deep 24\,$\micro$m catalogs associated with photometric redshift and information provided by the stacking of these sources into the deep SPIRE maps of the GOODS-N and COSMOS fields. Finally, by integrating all these counts, we studied the contribution of the galaxies to the CIB as a function of their flux density and redshift.
}{
Through stacking, we managed to reconstruct the source counts per redshift slice down to $\sim$2\,mJy in the three SPIRE bands, which lies about a factor 10 below the 5$\sigma$ confusion limit. Our measurements place tight constraints on source population models. None of the pre-existing models are able to reproduce our results at better than 3-$\sigma$. Finally, we extrapolate our counts to zero flux density in order to derive an estimate of the total contribution of galaxies to the CIB, finding 10.1$_{-2.3}^{+2.6}$\,nW\,m$^{-2}$\,sr$^{-1}$, 6.5$_{-1.6}^{+1.7}$\,nW\,m$^{-2}$\,sr$^{-1}$, and 2.8$_{-0.8}^{+0.9}$\,nW\,m$^{-2}$\,sr$^{-1}$ at 250\,$\micro$m, 350\,$\micro$m, and 500\,$\micro$m, respectively. These values agree well with FIRAS absolute measurements, suggesting our number counts and their extrapolation are sufficient to explain the CIB. We find that half of the CIB is emitted at $z=$1.04, 1.20, and 1.25, respectively. Finally, combining our results with other works, we estimate the energy budget contained in the CIB between 8\,$\micro$m and 1000\,$\micro$m: $26_{-3}^{+7}$\,nW\,m$^{-2}$\,sr$^{-1}$.
}{}


\titlerunning{Deep number counts at 250\,$\mu$m, 350\,$\mu$m and 500\,$\micro$m and CIB build-up.}

\authorrunning{B\'ethermin et al.}

\begin{document}

\maketitle

\section{Introduction}

About half of the relic energy arising from the emission of galaxies, which we refer to as the Extragalactic Background Light (EBL), is contained in the cosmic infrared background (CIB), which lies between 8\,$\micro$m and 1000\,$\micro$m, and peaks at around 150\,$\micro$m \citep{Hauser2001,Dole2006}. The first absolute measurements of the CIB were performed in the nineties with the Far-Infrared Absolute Spectrophotometer (FIRAS; \citealt{Puget1996,Fixsen1998,Lagache1999}) and the Diffuse Infrared Background Experiment (DIRBE; \citealt{Hauser1998}) onboard the \textit{COsmic Background Explorer} (\textit{COBE}). The far-infrared emission from galaxies is mainly due to dust heated by ultraviolet photons re-radiate in the infrared. A small fraction of these far-infrared emission ($\sim$15\%) are due to accretion processes \citep{Alexander2005,Jauzac2011}. The CIB thus primarily gives a budget of infrared photons emitted by star-formation processes.\\

More recently, deep number counts (flux density distributions of infrared sources) have been measured in the mid- and far-infrared domain, thanks to the sensitivity of the \textit{Spitzer} \citep{Werner2004} and \textit{Herschel}\footnote{Herschel is an ESA space observatory with science instruments provided by European-led Principle Investigator
consortia and with important participation from NASA.} \citep{Pilbratt2010} space telescopes. They exhibit power-law behavior at the faint end, which can be extrapolated to zero flux density in order to estimate the contribution of all the galaxies to the CIB (e.g. \citealt{Papovich2004} at 24~$\mu$m with \textit{Spitzer}/MIPS, \citealt{Bethermin2010a} at 24~$\mu$m, 70~$\mu$m, and 160~$\mu$m with \textit{Spitzer}/MIPS, \citealt{Berta2011} at 70~$\mu$m, 100~$\mu$m, and 160~$\mu$m with \textit{Herschel}/PACS). These estimations of the total CIB agree with the absolute measurements performed by \textit{COBE}, suggesting the CIB is now explained shortward of 160~$\mu$m. At longer wavelengths, due to source confusion \citep{Dole2003,Nguyen2010}, the \textit{Herschel}/SPIRE instrument \citep{Griffin2010} can directly resolve only 20\%, 12\%, and 6\% of the CIB at 250\,$\micro$m, 350\,$\micro$m, and 500\,$\micro$m, respectively \citep{Oliver2010}. Due to the limited depth of these confusion-limited observations, the break and the power-law behavior of the counts at faint flux density cannot be seen. It is thus necessary to use statistical tools like $P(D)$ analysis\footnote{$P(D)$ analysis is a statistical method used to estimate the number counts in a field from the pixel histogram of an extragalactic map.} \citep{Condon1974,Patanchon2009} or stacking \citep{Dole2006,Marsden2009} to study the origins of the sub-mm part of the CIB.\\ 

Using a P(D) analysis, \citet{Patanchon2009} produced deep counts from the Balloon-borne Large-Aperture Submillimeter Telescope (BLAST, \citealt{Pascale2008,Devlin2009}) data. They were only able to constrain one data point below 100\,mJy at 250\,$\micro$m, and were not sensitive to more subtle features in the shape of the counts. Using a stacking analysis, \citet{Bethermin2010b} managed to detect the peak of the Euclidian-normalized counts at 250\,$\micro$m in the BLAST data, but not at longer wavelengths. Using a $P(D)$ analysis on SPIRE data, \citet{Glenn2010} managed to clearly detect this peak at 250\,$\micro$m and 350\,$\micro$m, but not at 500\,$\micro$m. In all these cases, the uncertainties are too large to reliably detect a power-law behavior at the faint end. \\

A stacking analysis of  SPIRE data similar to that performed on BLAST data by \citet{Bethermin2010b} could significantly reduce the uncertainties and provide more precise information on the sources which make up the CIB. \citet{Le_Floch2009} and \citet{Berta2011} also showed that counts per redshift slice are strong constraints for galaxy evolution models (e.g. \citealt{Le_Borgne2009}, \citealt{Valiante2009}, \citealt{Marsden2010}, \citealt{Bethermin2011}, \citealt{Gruppioni2011}, \citealt{Rahmati2011}). Lastly, unlike a P(D) analysis, stacking allows us to measure directly the counts in redshift slices, but requires a prior catalog. Thus, here we perform a stacking analysis in the SPIRE bands, in the COSMOS and GOODS-N fields to produce deep counts per redshift slice in SPIRE bands, combining the \textit{Herschel} Multi-tiered Extragalactic Survey (HerMES)\footnote{hermes.sussex.ac.uk} data \citep{Oliver2011} and the ancillary data.\\

The paper is organized as follows. In Sect.~\ref{section:data}, we present the different data sets used in our analysis. We then introduce the method used to measure the number counts of resolved sources (Sect.~\ref{sect:prior_method}) and another method based on stacking to reconstruct the number counts below the confusion limit (Sect.~\ref{sect:stacking}). In Sect.~\ref{sect:uncertainties}, we detail the estimation of the statistical uncertainties. Sect.~\ref{sect:simulation} presents a end-to-end simulation used to check the accuracy of our method. In Sect~\ref{sect:number_counts}, we interpret our number counts and compare them with previous measurements and models of galaxy evolution. The same thing is done in Sect.~\ref{sect:redshift_distributions} for the redshift distributions. In Sect.~\ref{sect:cib}, we derive constraints on the CIB level and its redshift distribution from our number counts. We finally discuss our results (Sect.~\ref{sect:discussion}) and conclude (Sect.~\ref{sect:conclusion}).

\section{Data}

\label{section:data}

\subsection{SPIRE maps at 250\,$\micro$m, 350\,$\micro$m and 500\,$\micro$m}

The SPIRE instrument \citep{Griffin2010} onboard the \textit{Herschel Space Observatory} \citep{Pilbratt2010} observed the COSMOS field as part of the \textit{Herschel} Multi-tiered Extragalactic Survey (HerMES) program \citet{Oliver2011}. The maps were built using an iterative map-making technique \citep{Levenson2010}. The full width at half maximum (FWHM) of the SPIRE beam \citep{Swinyard2010} is 18.1\arcsec, 24.9\arcsec, and 36.6\arcsec\ at 250\,$\micro$m, 350\,$\micro$m, and 500\,$\micro$m, respectively. The typical instrumental noise is 1.6, 1.3, and 1.9\,mJy\,beam$^{-1}$ in COSMOS (1.6, 1.3, and 2.0\,mJy\,beam$^{-1}$ in GOODS-N) and the 1-$\sigma$ confusion noise is 5.8, 6.3 and 6.8\,mJy\,beam$^{-1}$ in the three wavebands \citep{Nguyen2010}. The maps are thus confusion limited. The absolute calibration uncertainties in point sources are estimated to be 7\% (Swinyard et al. 2010, updated in the SPIRE Observers' Manual\footnote{http://herschel.esac.esa.int/Docs/SPIRE/pdf/spire\_om.pdf}).\\

\subsection{Ancillary data in COSMOS}

\label{sect:cosmos_catalog}

Deep 24\,$\micro$m imaging of the COSMOS field was performed by the \textit{Spitzer Space Telescope} (S-COSMOS, \citealt{Sanders2007}). The associated catalog reaches 90\% completeness at 80\,$\micro$Jy \citep{Le_Floch2009}. This catalog was matched with photometric redshifts of \citet{Ilbert2009}. Due to the high density of optical sources compared with the size of the MIPS beam, the cross-identification can be ambiguous in many cases. An intermediate matching was thus performed with the $K$ and IRAC bands where the source density is smaller, which helps to discriminate between several optical counterparts in a MIPS beam \citep{Le_Floch2009}.\\

In this paper, we use an updated version of the photometric redshift catalog of \citet{Ilbert2009} (v1.8). This version uses new deep $H$-band data. However, this catalog is not optimized for AGN. For the sources detected by XMM-\textit{Newton} we instead use the photometric redshifts of \citet{Salvato2009}, estimated with a technique specific to AGN. In addition, 10~000 sources have spectroscopic redshifts provided by the S-COSMOS team \citet{Lilly2007}, which, where available, are used instead of the photometric redshifts. Details of the updated COSMOS $S_{24}+z$ catalog will be given in Le Floc'h et al. (in prep.). In this new version, 96\% of the 27\,811 $S_{24}>80\,\mu$Jy sources have redshifts (9.7\% of them are spectroscopic).\\

\subsection{Ancillary data in GOODS-N}

In the GOODS-N field, we use the 24\,$\micro$m catalog of \citet{Magnelli2011}. This catalog was built using the IRAC catalog at 3.6\,$\micro$m as a prior, and has an estimated 3-$\sigma$ depth of 20~$\mu$Jy, but at this depth the completeness is only $\sim$50\%. The stacking of an incomplete catalog can bias the results (\citealt{Bethermin2010b}; Heinis et al. in prep., Vieira et al. in prep.). According to simulations, cutting at 80\% completeness results in smaller bias. We thus cut the catalog at 30\,$\mu$Jy (80\% completeness limit) to have a more complete and reliable sample. These sources were matched with the photometric redshifts of \citet{Eales2010} (97.4\% of the 2791 24\,$\mu$m sources are associated with a redshift). The data fusion of these two catalogs will be explained in more detail in Vaccari et al. (in prep.).\\

\section{Measuring the statistical properties of the resolved sources}

\label{sect:prior_method}

In order to build counts per redshift slice and redshift distributions of the sources selected by their SPIRE flux densities, we require catalogs containing SPIRE flux densities and redshifts. The redshift catalogs are built from optical and near-infrared catalogs. We start from catalogs of the 24\,$\micro$m sources which have optical counterparts and thus photometric redshifts. Due to the large beam of SPIRE, it is not trivial to identify the MIPS 24\,$\micro$m counterpart for a given SPIRE source. To avoid this problem, we directly measure the SPIRE flux denisty of the 24\,$\micro$m sources in the maps by PSF-fitting assuming a known position \citep{Bethermin2010b,Chapin2010,Roseboom2010}. Sect.~\ref{sect:dropouts} and \ref{sect:number_counts} discuss the relevance of this choice of prior.\\

The GOODS-N field, being much smaller than COSMOS, has little impact on the statistical uncertainties (using GOODS-N+COSMOS reduces the uncertainties of 0.2\% compared to COSMOS only). The inclusion of GOODS-N introduces heterogeneity to the 24 um catalogs, which are built using different methods between the two fields. For this reason, we used only the COSMOS field in the following section.\\

\subsection{Source extraction}

We use the \textsc{fastphot} PSF-fitting routine, described in \citet{Bethermin2010b}, to fit the following model to the SPIRE map: 
\begin{equation}
m = \sum_{k=1}^{N_s} S_k \times b_{x_k,y_k}+\mu,
\end{equation}
where $m$ is the map, $N_s$ the number of sources, $S_\textrm{k}$ the SPIRE flux density of the $k$-th source, $b_{x_k,y_k}$ a point spread function (PSF) centered on the position of the k-th source ($x_k,y_k$), and $\mu$ a constant background. The catalog of positions used as input to \textsc{fastphot} is discussed below. The free parameters fit by \textsc{fastphot} are the SPIRE fluxes of sources in the prior list $S_k$ and the level of the constant background $\mu$. We used the PSF based on the Neptune scan from \citet{Glenn2010} \footnote{Beam data are also available from the Herschel Science Centre at ftp://ftp.sciops.esa.int/pub/hsc-calibration/SPIRE/PHOT/Beams}. The map is not fit in one pass, but split into 100$\times$100 pixel regions (the pixel sizes are 6.0\arcsec, 8.3\arcsec, and 12\arcsec at 250\,$\micro$m, 350\,$\micro$m, and 500\,$\micro$m, respectively). Each region was fit independently. To limit the edge effects, we also fit simultaneously an additional region of 20\,pixels around each 100$\times$100 region. The positions of the sources in both the central and additional regions are used by \textsc{fastphot}, but we keep in the final catalog only the photometry of the sources in the central region. The signal at 20\,pixels ($\sim$6 times the FWHM) from the center of a source is negligible. A source outside of the additional region cannot thus significantly affect the photometry in the central region.\\

The \textsc{fastphot} routine suffers some instabilities when two sources are too close to one another. We thus do not use the position of all the 24~$\mu$m sources in \textsc{fastphot}. For several redshift and 24\,$\micro$m flux density slices, we estimated the mean color by stacking (see Sect.~\ref{sect:stacking_method}). We then use these mean colors to estimate the flux density of each source in the SPIRE bands. A 24\,$\micro$m source is included in the position list of \textsc{fastphot} only if it has the highest estimated SPIRE flux density in a 0.5$\times$FWHM radius. This process was therefore performed independently in each band. Some sources with unusually high sub-mm/mid-infrared colors could be missed by this method, but there are few objects of this kind (see Sect.~\ref{sect:dropouts}).\\

To avoid unphysical negative flux densities for faint sources lying on negative fluctuations of the noise, we run \textsc{fastphot} iteratively, removing from the position list the sources with negative flux densities at each iteration. Removing a source from the input catalog is equivalent to assuming its flux density is zero, which is the most probable value in this case.\\

\subsection{Estimating photometric noise}

\label{sect:noise}

We estimate the photometric noise from the standard deviation of the \textsc{fastphot} residual map, finding the values 4.6, 5.5 and 5.1\,mJy at 250\,$\micro$m, 350\,$\micro$m, and 500\,$\micro$m, respectively. These values are about 20\% lower than the combination the 1-$\sigma$ confusion noise measured by \citet{Nguyen2010} and the instrumental noise (6.0, 6.4 and 7.0\,mJy). Our method is thus more efficient than a naive blind extraction. We chose to cut our statistical analysis at 20\,mJy in the three SPIRE passbands, which corresponds to about 4-$\sigma$.\\

\begin{figure}
\centering
\includegraphics{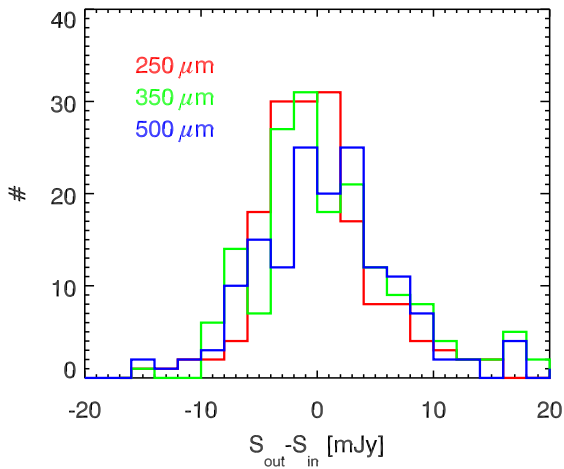}
\caption{\label{fig:artificial_sources}Simulation of the photometric uncertainties: histogram of the difference between the recovered and the input flux density of the artificial sources injected in the real SPIRE map and re-extracted with \textsc{fastphot}.}
\end{figure}

In order to cross-check our estimate of the photometric noise, we inject 200 artificial point sources in the real SPIRE map and add them in the input position list of \textsc{fastphot}. We add a random shift drawn from a 2D Gaussian with $\sigma=2\arcsec$ to the source position in order to simulate the astrometric uncertainties of the real catalog. We then rerun \textsc{fastphot} and compare the input and output flux densities. Fig.~\ref{fig:artificial_sources} shows the histogram of the difference between the recovered and input flux densities. We found a 1-$\sigma$ photometric noise of 3.9, 5.2 and 5.1\,mJy at 250\,$\micro$m, 350\,$\micro$m, and 500\,$\micro$m, respectively. The values are similar to those estimated from the residual map. Comparing the two sets of values, we can estimate an error of about 20\% on the photometric noise.\\ 

\subsection{24\,$\micro$m Dropouts}

\label{sect:dropouts}

A fundamental limitation of our model is that it is not sensitive to any population of sources that are faint at 24 microns but bright in the SPIRE passbands (24~$\mu$m dropouts).  No such population is known or theoretically predicted, except possibly at very high redshifts, but the possibility remains that nature has been more inventive than we have.  In this section we attempt to test whether there is any evidence for such sources, and do not find any.\\

First, we study the residual SPIRE maps after removing all of the sources extracted with FASTPHOT to estimate the number of sources missed by the extraction.  The density of remaining sources is quite small so that confusion is not a problem.  We then search for additional sources by looking for peaks in the beam smoothed residual map to a depth of 20~mJy, and find that the fraction\footnote{The fraction of dropouts is defined as $\frac{N_d}{N_d+N_p}$, where $N_d$ is the number of sources brighter than 20~mJy found in the residual map and $N_p$ the number of sources brighter than 20~mJy extracted by \textsc{fastphot} in the normal map.} of such possible sources is only 1.3\%, 0.7\%, and 0.6\% at 250 $\mu$m, 350 $\mu$m, and 500 $\mu$m, respectively. Next, we have compared our prior catalog with the source list from the blindly-extracted HerMES catalog \citep{Smith2012}, which is limited to sources brighter than 20 mJy. The fraction of sources in the blind catalog without counterpart in the prior catalog strongly depends on the choice of matching radius. For a narrow radius of 0.5 FWHM, we obtained 3.2\%, 2.6\%, and 0.6\% whereas for a large radius of 1 FWHM, we obtained 0.6\%, 0.3\%, and 0.0\%, at 250 $\mu$m, 350 $\mu$m, and 500 $\mu$m, respectively. However, with the narrow radius, we miss some sources due to astrometric uncertainties, and with the large radius, we possible have a contamination by neighboring sources. It is expected that the fraction of dropouts decreases with the flux density. Nevertheless, this behavior is hard to constrain because of the small number of bright sources. Note however that the fraction of dropouts at the flux density limit is very close to the values obtained for the full sample because of the steep slope of the counts. Thus, we conclude that our catalog, based on the 24\,$\micro$m prior, is very close to complete above 20~mJy. From a blind extracted catalog of H-GOODS data, \citet{Magdis2011} estimated the fraction of 24~$\mu$m dropouts for H-GOODS fields and shallower fields. They predict a dropout faction smaller than 2\% in the COSMOS field.\\

Finally, we can compare our number counts measurement to other analyses that did not make use of a 24~$\mu$m prior.  We find good agreement with the blind extractions of \citet{Oliver2010} and \citet{Clements2010}, but note that these analyses required significant model corrections for Eddington bias and confusion.  We also find good agreement with the P(D) analysis of the SPIRE maps by \citet{Glenn2010}, which, by construction, is not affected by either issue.  We take these comparisons as a strong indication that we have not missed a statistically significant population, at least in terms of the redshift integrated number counts.  However, we must acknowledge that the fraction of dropouts could evolve with redshift, and in particular the high redshift bins may be less complete.\\

\subsection{Correction of the biases}

\label{sect:res_corr}

\begin{figure}
\centering
\includegraphics{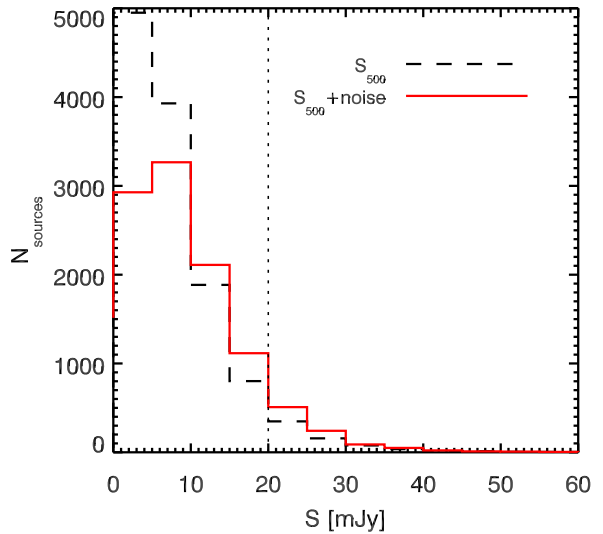}
\caption{\label{fig:edd_boost} Effect of the photometric uncertainties in the flux density distribution at 500\,$\micro$m. \textit{Black dashed line}: distribution of the flux density measured at the position of the 24\,$\micro$m sources. \textit{Red solid line}: the same distribution after adding a 5.1\,mJy random Gaussian noise to each measured flux density. Due to photometric noise, some sources have a negative flux density (put to zero in our iterative algorithm) and are not represented here. \textit{Black dotted line}: flux density cut used in our analysis (20~mJy).}
\end{figure}

Intuitively, the simplest way to compute the source counts is to measure the number of sources in a flux density bin and divide it by the width of the bin and the surface area of the field. However, due to photometric noise, this estimate is biased. In fact, for a prior-based extraction, we do not have a flux boosting phenomena (which appears at low signal to noise ratio for blind extraction, because the completeness is higher for sources lying on positive fluctuations of the noise, as discussed e.g. in \citealt{Bethermin2010b}), but another statistical effect, Eddington bias, biases the counts measurement, as illustrated by Fig.~\ref{fig:edd_boost}. The black dashed line shows the distribution of the 500~$\mu$m flux densities measured at the prior positions. We will assume this distribution is close to the real one, and will somewhat arbitrarily refer to it as initial distribution. The red line shows the same distribution, but adding a 1-$\sigma$ 5.1~mJy Gaussian error on the flux density of each source, called measured distribution. At bright flux density (S$_{500}>20$\,mJy), we can observe an excess in the measured distribution compared to the initial one.\\

To correct this bias, we use a Monte Carlo (MC) method as in \citet{Bethermin2010b}. We compute 1000 realizations of the bias in each flux density (regular in logarithm from 20~mJy) and redshift ($0<z<0.5$, $0.5<z<1$, $1<z<2$, and $z>2$) bin, and use them to compute the mean correction and its uncertainty:
\begin{itemize}
\item We start from the measured distribution of the 250\,$\mu$m, 350\,$\mu$m, or 500\,$\mu$m flux density of the 24~$\mu$m sources in the prior list of \textsc{fastphot} in a given redshift bin, and assume it is close to the initial distribution. This last hypothesis is a significant approximation, but the selection function of the procedure used to construct the prior catalog is too complex to be modeled without introducing strong assumptions about galaxy evolution.
\item We draw with replacement N sources in the initial sample, where N is the number of sources in the initial sample. This bootstrap step is used to take into account the sample variance on the initial flux density distribution.
\item We add a Gaussian random photometric noise to the flux density of each source. We use the values of the noise found in Sect.~\ref{sect:noise} plus a 20\% systematic shift (different at each iteration of the MC procedure), which takes into account the systematic uncertainty on the determination of the noise.
\item We compute the bias on the counts dividing the counts from the drawn sample before and after adding the photometric noise.
\end{itemize}
Table \ref{fig:rescf} shows the corrective factor in various flux density and redshift bins and at various wavelengths. This correction can reach 40\% in the fainter flux  density bins and decreases at brighter flux densities.\\

Similar corrections are applied to the redshift distributions. However, in addition, we apply a random error to the redshift based on the uncertainties provided in the photo-z catalog (but without taking into account the catastrophic outliers) during the MC procedure. In this case, there is only one flux density bin ($>$20~mJy).\\

\section{Measuring the statistical properties of sources below the confusion limit}

\label{sect:stacking}

Due to source confusion, SPIRE cannot resolve the bulk of the CIB into individual sources \citep{Nguyen2010, Oliver2010}. Nevertheless, about 80\% of the CIB is resolved at 24\,$\micro$m \citep{Papovich2004,Bethermin2010a}. We thus perform a stacking analysis using the 24\,$\micro$m prior to probe fainter populations and resolve a larger fraction of the sub-mm CIB.\\

\subsection{Stacking method}

\label{sect:stacking_method}

Stacking is a statistical method which allows us to measure the mean flux density of a population of sources selected at another wavelength, but which are too faint to be detected individually at the working wavelength. Several methods can be used (e.g. \citealt{Dole2006}, \citealt{Marsden2009}, see the discussion in Vieira et al. in prep.). We use the following method (also used in Vieira et al. in prep.): we first subtract the mean of the SPIRE map in the region covered by the 24\,$\micro$m observations. We then compute the mean signal in pixels which has a source centered on them. This provides the mean flux density of the population, because the SPIRE maps are in Jy\,beam$^{-1}$. Vieira et al. (in prep.) showed that this method is more accurate in a confusion-limited case than PSF-fitting on a stacked image. The uncertainties are estimated using a bootstrap method \citep{Jauzac2011}.\\

\begin{figure}
\centering
\includegraphics{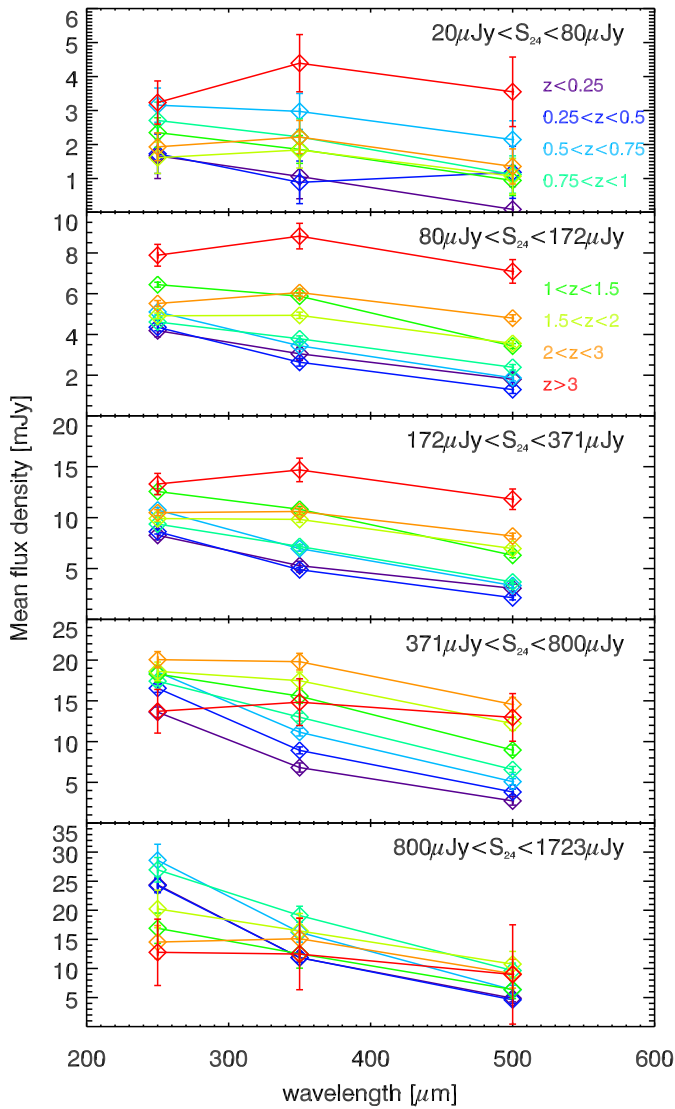}
\caption{\label{fig:mean_flux} Mean flux density measured by stacking as a function of wavelength. The various redshift bins are represented using various colors. Each panel corresponds to each 24~$\mu$m flux density bins. The error bars are estimated with a bootstrap method.}
\end{figure}

Due to the large number of sources in COSMOS, we can split our 24\,$\micro$m sample into eight redshift bins ($0<z<0.25$, $0.25<z<0.5$, $0.5<z<0.75$, $0.75<z<1$, $1<z<1.5$, $1.5<z<2$, $2<z<3$, and $z>3$) and logarithmic flux density slices ($80 \, \mu$Jy$<S_{24}<172 \, \mu$Jy, $172 \, \mu$Jy$<S_{24}<371 \, \mu$Jy, $371 \, \mu$Jy$<S_{24}<800 \, \mu$Jy, and $800 \, \mu$Jy$<S_{24}<1723 \, \mu$Jy). In GOODS-N, we use the same redshift slices, but a single flux density slice ($30 \, \mu$Jy$<S_{24}<80 \, \mu$Jy). This choice of the number of bins was done to have a compromise between a fine grids in 24\,$\mu$m flux density and redshift, but also a reasonable number of sources to stack in each bins to obtain a good signal-to-noise ratio. We stack the sources in each bin to compute their mean flux density in the three SPIRE bands. Fig.~\ref{fig:mean_flux} shows the mean flux density as a function of wavelength, which, as expected, decreases rapidly in low redshift bins and peaks between 350~$\mu$m and 500~$\mu$m in the $z>3$ bin. The mean color in each bin is computed by dividing the mean SPIRE flux density by the mean 24\,$\mu$m flux density.

\subsection{Scatter of the photometric properties of the stacked populations}

\label{sect:scatter}

The uncertainties given by the bootstrap method, $\sigma_\textrm{boot}$, are 
\begin{equation}
\label{eq:boostrap}
\sigma_{\textrm{boot}} = \frac{\sqrt{\sigma_{\textrm{instr}}^2+\sigma_{\textrm{conf}}^2+\sigma_{\textrm{pop}}^2}}{\sqrt{N_{\textrm{stack}}}},
\end{equation}
where $\sigma_{instr}$ is the instrumental noise, $\sigma_{\rm conf}$ the confusion noise, $\sigma_{\rm pop}$ is the standard deviation of the flux density of the population, and $N_{\rm stack}$ the number of stacked sources. The quantity $\sqrt{\sigma_{\rm instr}^2+\sigma_{\rm conf}^2}$ can be estimated from the standard deviation of the map, allowing us to deduce $\sigma_{\rm pop}$ from our bootstrap analysis.\\

\begin{figure}
\centering
\includegraphics{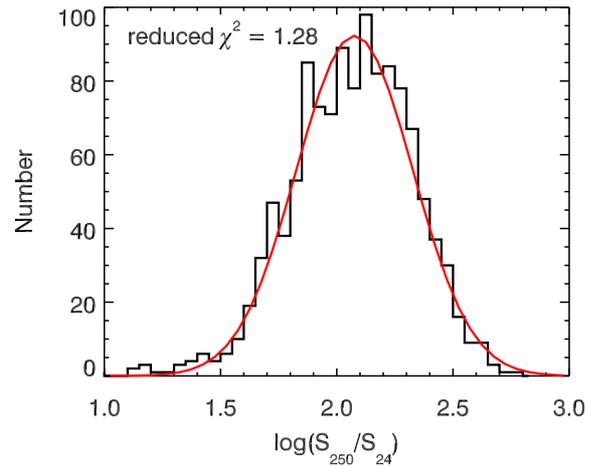}
\caption{\label{fig:color_distribution} \textit{Black histogram}: distribution of the logarithm of the $S_{250}/S_{24}$ colors of the resolved sources ($S_{250}>20\,$mJy) in the 1$<$z$<$1.5 redshift bin. \textit{Red line}: fit of the histogram by a Gaussian.}
\end{figure}

\begin{figure}
\centering
\includegraphics{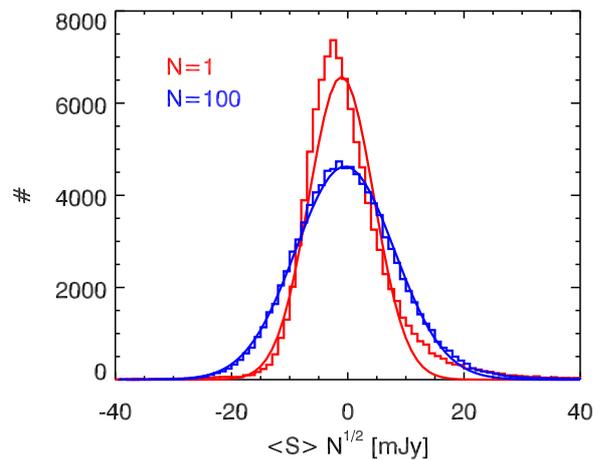}
\caption{\label{fig:pd_to_gaussian} \textit{Red histogram}: Pixel histogram of the 250~$\mu$m SPIRE map in COSMOS. \textit{Blue histogram}: Histogram of the mean signal in 100 pixels taken randomly in 100\,000 realizations. \textit{Red and blue lines}: Gaussian fit of the previous histograms.}
\end{figure}

While this formula is true for a Gaussian distribution, we note that the distribution of colors of the sources are probably better described by a log-normal distribution. Fig.~\ref{fig:color_distribution} shows the distribution of the logarithm of the $S_{250}/S_{24}$ colors of the resolved sources ($S_{250}>20\,$mJy) in the $1<z<1.5$ redshift bin (this redshift bin was chosen because it has the larger number of sources). The confusion noise is also non Gaussian \citep{Glenn2010}. Nevertheless, due to central limit theorem, these distributions of the mean flux density tend to be Gaussian if a sufficient number of sources are stacked. Fig.~\ref{fig:pd_to_gaussian} illustrates this property. The red histogram is the pixel histogram of the 250\,$\mu$m SPIRE map. It is not Gaussian, because the confusion noise is not. The blue one is the distribution of the mean signal in 100 pixels taken randomly in 100\,000 realizations (typically the effect of the instrumental and confusion noise on a stack of 100~sources). This histogram is much closer to a Gaussian. The same thing happens for the color scatter term. The Gaussian approximation is thus very relevant here.\\

The scatter on the $S_{\rm SPIRE}/S_{24}$ color can be estimated by dividing $\sigma_{\rm pop}$ by the mean flux density of the population. We do not detect a significant evolution of this scatter with redshift, wavelength or 24\,$\micro$m flux density. We use the median and the standard deviation of the values found in the different redshift and 24\,$\micro$m flux density bins, and find a scatter $\sigma_{\rm color} = \sigma_{\rm pop} / <S_{\rm SPIRE}>$ of 68$\pm$35\% (in linear units). This agrees with the value of 62\% found for the resolved sources (see Fig.~\ref{fig:color_distribution}).\\

\subsection{Does the clustering of sources introduce a bias?}

\label{sect:clusbias}

The simplest stacking method assumes implicitly that the sources in the map are not clustered, but this has been shown to be unrealistic and must be accounted for \citep{Bethermin2010b,Viero2011,Penner2011}. We have performed several tests in the COSMOS field to estimate the bias due to clustering, which we now describe.\\

\subsubsection{Method A: convolution of the 24\,$\micro$m map with the SPIRE beam}

A simple way to estimate the bias due to clustering is to convolve the 24\,$\micro$m map with a Gaussian kernel to obtain a 24\,$\micro$m map with a Gaussian PSF of the same FWHM as the SPIRE map \citet{Oliver2010b}. To match resolutions, we use a Gaussian kernel with beamsize $\sigma_{\rm kernel} =\sqrt{\sigma_{\rm SPIRE}^2-\sigma_{\rm MIPS}^2}$. We measure the mean flux density of the 24\,$\micro$m sources by stacking the 24~$\mu$m catalog on this convolved 24\,$\micro$m map. The bias due to clustering is estimated by comparing the mean flux density measured by stacking with the mean flux density estimated from the 24~$\mu$m catalog. We find biases of 5$\pm$2\%, 11$\pm$2\%, and 20$\pm$5\% at 250\,$\micro$m, 350\,$\micro$m, and 500\,$\micro$m, respectively. This method is equivalent to building a simulated map assuming a single color $C$ for all the 24~$\mu$m objects (including the ones below the detection limit at 24~$\mu$m), measuring the mean flux density of the selected population by stacking on this convolved map, and comparing it with the mean flux density coming from the catalog ($<S_{24}>\times C$). The same color factor $C$ is present in the mean stacked flux measured by stacking in the convolved map and in the mean flux coming from the catalog. It thus disappears when we compute the relative difference between these two quantities. We thus take $C=1$ for simplicity. As expected, the bias due to clustering increases with the size of the beam. This estimate is exact only if the $S_{SPIRE}/S_{24}$ color is constant, or if the properties of the angular clustering do not evolve with the color of the sources (and thus the redshift). These assumptions are not going to be exactly met, so that next we use another method to cross-check this estimate.\\

\subsubsection{Method B: simulation based on mean colors measured by stacking}

In Sect.~\ref{sect:stacking_method}, the $S_{\rm SPIRE}/S_{24}$ mean color as a function of the 24\,$\micro$m flux density and redshift were measured by stacking. We use these mean colors and the scatter measured in Sect.~\ref{sect:scatter} to generate mock SPIRE flux densities for the sources in the $S_{24}+z$ catalog, and then build a simulated map of the COSMOS field using the position given in the $S_{24}+z$ catalog, the estimated SPIRE flux density, and the SPIRE PSF.  Random Gaussian noise was added following the noise map of the real data. We then stacked all the 24\,$\micro$m sources, and compared the mean flux density measured by stacking in the simulated map and the mean flux density in the mock catalog. We find a bias of 7.0$\pm$0.9\%, 10.4$\pm$ 0.7\%, and 20.6$\pm$1.2\% at 250\,$\micro$m, 350\,$\micro$m, and 500\,$\micro$m, respectively, in agreement with the values provided by method A. The main drawback of this method is that any bias due to sources undetected at 24 microns is not modeled.\\

We have also stacked sub-samples selected in redshift and/or in 24\,$\micro$m flux density. The bias tends to slightly decrease with the redshift and the 24\,$\micro$m flux density cut. Nevertheless, this evolution is small (below 3\%), and the significance is smaller than 3-$\sigma$. We thus chose to neglect it, and assume a single value for the bias due to the clustering.\\

\subsubsection{Method C: fitting the profile of the stacked image}

\begin{figure}
\centering
\includegraphics{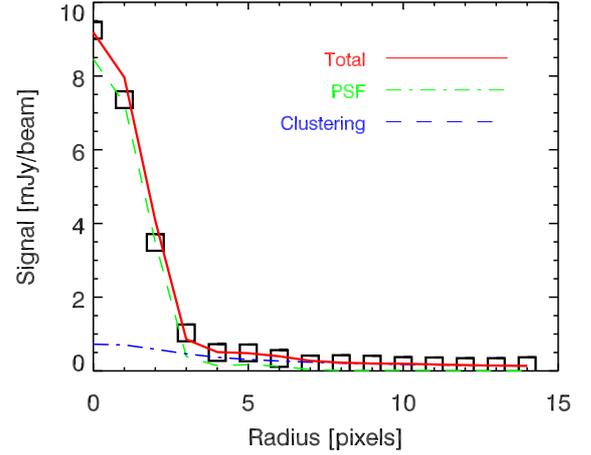}
\caption{\label{fig:clusfit} Radial profile of the stacked image at 250\,$\micro$m of all the 27\,811 $S_{24}>80\,\mu$Jy sources in COSMOS. \textit{Black squares}: measurements. \textit{Red solid line}: best fit. \textit{Green dot-dashed line}: contribution of the PSF. \textit{Blue dashed line}: contribution of the clustering. The error bars are too small to be represented. The pixel size is 6".}
\end{figure}

For method C, we follow the \citet{Dole2006} method to produce our stacked images.  In the presence of clustering, this image can be fit by the following function (\citealt{Bethermin2010b}; Henis et al. in prep.):
\begin{equation}
M = \alpha \times b + \beta \times \left ( \frac{b \ast w}{\textrm{max}(b \ast w)} \right ),
\end{equation}
where $M$ is the stacked image, $w$ the auto-correlation function (ACF), $\ast$ the convolution product, and $b$ the beam function. The PSF is normalized to unity at the center to match the per-beam normalization of the maps. $\alpha$ and $\beta$ are free parameters in the fit. The results of the fit are plotted in Fig.~\ref{fig:clusfit}. In order to estimate the uncertainties, the fit was performed on 1000 bootstrap samples. If we measure the photometry in the central pixel of the PSF, the bias due to clustering is $\beta/\alpha$. We found 7.7$\pm$0.5\%, 10.3$\pm$0.8\%, and 19.1$\pm$1.8\% at 250\,$\micro$m, 350\,$\micro$m, and 500\,$\micro$m, respectively. The uncertainty here is the standard deviation of the values found for the different bootstrap samples. As expected, we also found that $\alpha$ and $\beta$ are significantly anti-correlated (correlation coefficients of $-0.46$, $-0.56$, and $-0.57$ at 250\,$\micro$m, 350\,$\micro$m, and 500\,$\micro$m, respectively). As was the case for method B, we do not detect any significant evolution of the bias with redshift.

\subsubsection{Correction of the bias due to clustering}

\begin{table}
\centering
\begin{tabular}{lrrr}
\hline
\hline
\noalign{\vspace{2pt}}
wavelength & \multicolumn{3}{c}{Bias due to clustering} \\
$\micro$m & Method A &  Method B & Method C\\
\noalign{\vspace{2pt}}
\hline
\noalign{\vspace{2pt}}
250 & 5$\pm$2\% & 7.0$\pm$0.9\% & 7.7$\pm$0.5\% \\
350 & 11$\pm$2\%& 10.4$\pm$ 0.7\% & 10.3$\pm$ 0.8\% \\
500 & 20$\pm$5\% & 20.6$\pm$1.2\% & 19.1$\pm$1.8\% \\
\noalign{\vspace{2pt}}
\hline
\end{tabular}
\caption{\label{tab:clusbias} Bias due to clustering as a function of the wavelength. These values are estimated with the methods presented in Sect.~\ref{sect:clusbias}.}
\end{table}

Table~\ref{tab:clusbias} summarizes our estimates of the bias due to clustering. Our three methods give similar results. To correct for the effects of the clustering, we divide our measured mean flux densities by the mean values of 1.07, 1.10, and 1.20 at 250\,$\micro$m, 350\,$\micro$m, and 500\,$\micro$m, respectively.\\

\subsection{Reconstruction of the SPIRE counts}

\label{sect:spire_reconstruction}

We can reconstruct the SPIRE counts using the information provided by the $S_{24}+z$ catalog, the mean color, and the standard deviation provided by the stacking analysis. In this analysis, we assume that the distribution of the SPIRE flux density for a given 24\,$\micro$m flux density is log-normal (see Fig.~\ref{fig:color_distribution} and Sect.~\ref{sect:scatter}). For a small scatter ($<<$1), the standard deviation of the logarithm of the flux density $\sigma_{\rm log-norm,\, color}$ can be computed from the standard deviation of the flux density $\sigma_{\rm color}$: $\sigma_{\rm log-norm,\, color} = \sigma_{\rm color}/\textrm{ln}(10)$. For larger scatter, this approximation is no longer valid. However, there is a bijective link between the following two pairs of parameters: the mean and the scatter of the color in linear units and the same thing in logarithmic units. We can thus deduce the two parameters of the log-normal distribution (mean and scatter) of the color from the linear mean and standard deviation measured by stacking.\\

We generate 1000 realizations of the SPIRE counts using the following recipe:
\begin{itemize}
\item We take randomly a value of the scatter (see Sect.~\ref{sect:scatter}). At each realization, we used a single value of the scatter for all the flux density and redshift bins.
\item In each flux density and redshift bin, we take randomly one value of the $S_{\rm SPIRE}/S_{24}$ color following the uncertainties (see Sect.~\ref{sect:stacking_method}). We obtain a relationship between $S_{24}$ and the color in each redshift slice interpolating between the centers of the 24\,$\mu$m bins.
\item We then compute the mean color of each source using the previous relationship.
\item For each source, we draw randomly a SPIRE flux density from its 24~$\mu$m flux, its color and the scatter on it. We assume a log-normal distribution.
\item We then compute the counts from the obtained SPIRE flux densities.
\end{itemize}
The final counts are computed taking the mean and the standard deviation of the different realizations.\\

\begin{figure}
\centering
\includegraphics{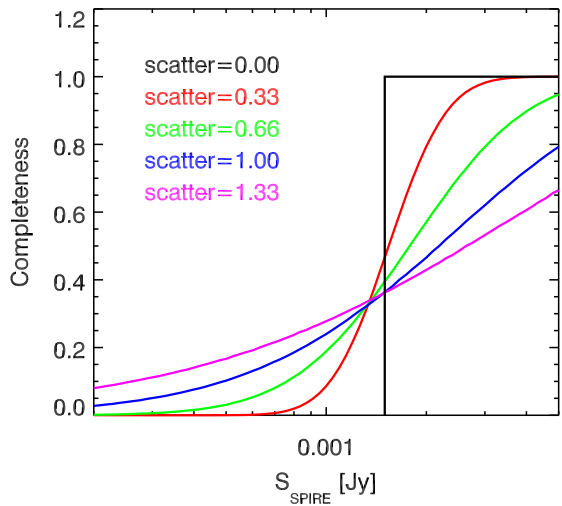}
\caption{\label{fig:stacking_completeness} Completeness of the SPIRE counts reconstructed by stacking in a simple case for various values of scatter on the colors. We assume power-law counts ($dN/dS\propto S^{-1.5}$), a mean SPIRE/24\,$\micro$m ratio, a log-normal scatter with different values and a flux density cut at 24\,$\micro$m of 30~$\mu$Jy. This figure is discussed in Sect.~\ref{sect:spire_reconstruction}.}
\end{figure}
 
Due to the flux density cut of the 24\,$\micro$m catalogs, the SPIRE simulated catalogs are not complete at the faint end. If there was a single color for all objects, the cut of the SPIRE catalog would be the SPIRE/24\,$\micro$m color multiplied by the flux density cut at 24\,$\micro$m. Above this limit, the catalog would be complete (statistically speaking), and there would be no sources below this limit. However, due to the scatter of the colors, this transition is smoother. We call the ratio between the reconstructed counts (taking into account the 24\,$\micro$m selection) and the input counts the completeness. \citet{Berta2011} used the \citet{Le_Borgne2009} model to estimate the completeness as a function of the far-infrared flux density. We have chosen to use a similar, but more empirical, method to estimate the completeness and correct for it. 
\begin{itemize}
\item We generate a mock 24\,$\micro$m catalog following power-law counts with a typical slope in $dN/dS \propto S^{-1.5}$ \citep{Bethermin2010a}.
\item We associate a SPIRE flux density with each source of the mock catalog using the real colors and scatters measured by stacking. The color of each source depends on its 24\,$\micro$m flux density and redshift.
\item In each SPIRE flux density bin, we compute the ratio between the total number of sources and the number of sources which are brighter than the 24\,$\micro$m flux density cut.
\end{itemize}
Several realizations of the colors and scatters are used to estimate the uncertainties in this correction. Fig.~\ref{fig:stacking_completeness} illustrates how the completeness values vary with the scatter of the colors in a simplified case, where we assume a single color for all the sources ($S_{\rm SPIRE}/S_{24}=50$) and a 24\,$\micro$m flux density cut of 30~$\mu$Jy. As expected, without scatter, the transition happens around 1.5\,mJy ($50\times0.03$), and the width of the transition increases with the scatter. Table \ref{stackcompgoodsn} and \ref{stackcompcosmos} provides the completeness corrections used in GOODS-N and COSMOS. We have cut our analysis in COSMOS at 6~mJy in the three SPIRE bands, because the completeness in the higher redshift bins is only $\sim$50\%. Below this limit, we used the GOODS-N field where the 24~$\mu$m catalog is deeper. Following the same criterion as in COSMOS, we cut our analysis at 2~mJy in the three bands. These cuts are slightly arbitrary, because the mean $S_{\rm SPIRE}/S_{24}$ color of the sources, and consequently the completeness, vary with redshift. Nevertheless, we use the same cuts for all redshifts in order to simplify the interpretation and the discussion.\\

The same type of analysis was performed to compute the redshift distribution of $S_{\rm 250, \, 350, \, or \,500}>6$\,mJy sources in COSMOS. In this case, there is only one flux density bin ($>6$\,mJy).

\section{Estimation of the statistical uncertainties}

\label{sect:uncertainties}

In Sect.~\ref{sect:prior_method} and \ref{sect:stacking}, we explained how we derived number counts and redshift distributions above and below the confusion limit. We also discussed the uncertainties in the corrections applied to our measurements. In this section, we explain how the field-to-field variance on our measurements is estimated and how we combine these uncertainties with the errors on the corrections.

\subsection{Sample variance}

Our study is based on only one or two fields depending on the flux density regime. The field to field variance cannot thus be easily estimated. We have used the same method based on the clustering of the sources as in \citet{Bethermin2010a}, which is briefly described here.

\subsubsection{Principle}

Spatially, sub-mm sources are not Poisson distributed \citep{Blain2004,Farrah2006,Cooray2010,Magliocchetti2011},  but clustered. The uncertainty, $\sigma_N$, on the number of sources in a given bin, $N$, is thus not $\sqrt{N}$. In large fields, this effect is not negligible, and the clustering of the sources must be taken into account \citep{Bethermin2010a}. The uncertainties in the clustered case are \citep{Wall2003}
\begin{equation}
\sigma_N = \sqrt{y N^2 + N},
\end{equation}
with
\begin{equation}
y = \frac{\int_{\rm field} \int_{\rm field} w(\theta) d\Omega_1 d\Omega_2}{\Omega^2},
\label{eq:ydef}
\end{equation}
where $w(\theta)$ is the auto-correlation function (ACF) and $\Omega$ the solid angle of the field. The effect of the clustering on the uncertainties depend only on the field (size and shape) and the ACF.\\

\subsubsection{Estimation of the auto-correlation function}

\label{sect:ls_est}

The purpose of this paper is not to study the clustering of sub-mm galaxies, but is just to compute, with a reasonable accuracy, its effect on uncertainties in the number counts. We measured the ACF of the resolved sources for the selection in redshift. This measurement is performed with the \citet{Landy1993} estimator:
\begin{equation}
w(\theta) = \frac{DD-2 \times DR+RR}{RR},
\end{equation}
where $DD$ is the number of pairs separated by an angle between $\theta-d\theta/2$ and $\theta+d\theta/2$ in the real catalog, $RR$ the number of pairs in a Poisson distributed catalog generated with the mask used for the source extraction, and $DR$ the number of pairs coming from a source in the real catalog and a source in the random catalog. The method used to quickly compute the number of pairs is described in Appendix~\ref{sect:acf_stacking}.\\

We fit our results with the following simple form \citep{Magliocchetti2011}:
\begin{equation}
w(\theta) = A \left ( \left ( \frac{\theta}{\rm 1~deg} \right ) ^{1-\gamma} - C \right ),
\label{eq:acf}
\end{equation}
where $\gamma$ is fixed at the standard value of 1.8. This simple form does not work at small scales ($<$2\arcmin), where the contribution of the clustering between the sources in the same dark matter halo is not negligible \citep[e.g.][]{Cooray2010}. We use only the scales larger than 2\arcmin in our analysis. The integral constraint $C$ is a factor taking into account the fact that \citet{Landy1993} estimator is biased for finite size survey. $C$ depends on the size and the shape of the field and the value of $\gamma$, and can be computed from
\begin{equation}
\label{eq:ic}
C = \frac{\int_{\rm field} \int_{\rm field} \left ( \frac{\theta}{\rm 1~deg} \right )^{1-\gamma} d\Omega_1 d\Omega_2}{\Omega^2}
\end{equation}
Combining Eq.~\ref{eq:ydef}, \ref{eq:acf}, and \ref{eq:ic}, we obtain:
\begin{equation}
y = A \times C.
\end{equation}
For our masks, $C = 1.72$ in the COSMOS field and 7.16 in GOODS-N. In order to compute the effect of the clustering on our error bars on the number counts, we thus have to estimate $A$ in the various redshift and flux bins used in our analysis.\\

\subsubsection{Uncertainties in the resolved number counts}

We measure the clustering of the resolved sources from the source list produced in Sect.~\ref{sect:prior_method}. If we use only the source in a given flux density and redshift bin, we do not obtain sufficient signal to noise. Therefore, we compute the ACF of all $S_{\rm SPIRE}>20$\,mJy sources in a single flux density bin, but four redshift bins (0$<$z$<$0.5, 0.5$<$z$<$1, 1$<$z$<$2, and z$>$2), and assume that the ACF does not evolve too much with flux density. We obtained very good fits in each redshift bins at each wavelengths (reduced $\chi^2<1.3$ in all bins). From these fits, we compute the value of the $y$ parameter and the sample variance on our measurements. The uncertainties coming from the sample variance are then combined with the uncertainties coming from the correction applied to the counts (see Sect.~\ref{sect:res_corr}).\\

Table \ref{tab:unctab_resolved} summarizes the relative contribution of the clustering term ($\sigma_{\rm clus} = \sqrt{yN^2}$) to the total sample variance $\left( \sigma_{\rm clus+poi} = \sqrt{\sigma_{\rm clus}^2 + \sigma_{\rm poi}^2} = \sqrt{yN^2+N} \right )$. This contribution is dominant in the low flux density bins ($\sim$85\%), and decreases in brighter flux density bins, where N is smaller. We also compared the sample variance with the uncertainties in the corrections. This last correction increases the uncertainties by less than 40\%. We are thus dominated by sample variance.\\

\subsubsection{Uncertaintes in the number counts measured by stacking}

The clustering of the SPIRE sources below the confusion limit ($<20$\,mJy) measured by stacking (see Sect.~\ref{sect:stacking}) cannot be measured directly. In our analysis, we started from the 24~$\mu$m population as a prior. We thus use the clustering of this population to compute the effect of the clustering on the uncertainties, assuming it is close to the one of the SPIRE faint sources. The ACF was measured in the same redshift bins as for the resolved sources. We then compute the sample variance, and combine it with the uncertainties coming from the completeness corrections, the colors, and the scatter.\\

Table \ref{tab:unctab_stacking_goodsn} and \ref{tab:unctab_stacking_cosmos} summarize the relative contribution of the clustering to the uncertainties. As for the resolved sources, the clustering term dominates the Poisson term in the sample variance. In contrast to resolved counts, the errors coming from the completeness correction and the uncertainties in the colors and the scatter dominate the sample variance. A possible bias, due to the assumption that the 24~$\mu$m and sub-confusion limit 250\,$\mu$m population have similar clustering properties, has therefore only a modest impact to our uncertainty budget.

\subsubsection{Uncertainties in the redshift distributions}

The ACF is difficult to measure in small redshift bins, because the number of sources is small and the signal-to-noise ratio is then poor. For this reason, we have estimated how the clustering evolves when we reduce the size of a redshift bin. To quantify this effect, we compute the ACF of the 24\,$\micro$m catalog (the signal for resolved SPIRE sources only is too low) in $1-dz/2<z<1+dz/2$ bins with $dz$ varying from 0.1 to 1. We find A $\propto dz^{-0.9}$. To compute the uncertainties in the redshift distribution, we thus use the ACF measured previously to compute the sample variance on the counts in large redshift bins ($0<z<0.5, 0.5<z<1, 1<z<2,$ and $z>2$), and apply the scaling relation $y \propto A \propto dz^{-0.9}$. We then derived the sample variance, and combine it with uncertainties in the correction factor for resolved counts and the ones coming from completeness corrections, colors, and scatter for counts measured by stacking.\\

\section{Validation on simulation}

\label{sect:simulation}

In order to check the accuracy of our methods used to measure the number counts, we have performed an end-to-end simulation. The clustering of the sources below the confusion limit is not well known, and its effect on stacking has been estimated in Sect~\ref{sect:clusbias} with three methods based on the data. We have thus chosen to use a simulation with a Poisson distribution of the sources, because it is easier to generate.\\

\subsection{Description of the simulation}

\label{sect:simu_desc}

Our simulation is based on \citet{Bethermin2011} model, which is a parametric model based on the \citet{Lagache2004} spectral energy distribution (SED) library (two populations: normal and starburst galaxies). This model uses a simple broken power-law evolutionary behavior of the characteristic luminosity and density of the luminosity function (LF). The free parameters of the model were determined by fitting observed counts (including the Herschel resolved counts published by \citet{Oliver2010}), LFs, and the CIB. The model has not been modified since the publication of the associated paper. Note that this model includes the contribution of strongly-lensed sources to the counts.\\

A mock catalog, containing the 24~$\mu$m,  250\,$\mu$m,  350\,$\mu$m,  and 500\,$\mu$m flux densities and the redshift of the sources, was generated following the model. We then build a map of the COSMOS field from this mock catalog, the SPIRE noise map, and the SPIRE PSF. We then redo all the analysis described Sect.~\ref{sect:prior_method} and \ref{sect:stacking} using the $S_{24}+z$ mock catalog and the simulated SPIRE maps. The SEDs of this model were not calibrated following the correlation between stellar mass and the star formation rate (roughly proportional to the infrared luminosity) and are thus not valid below 8~$\mu$m rest-frame. We thus cannot use this simulation at redshifts larger than 2.\\

\subsection{Results}

\begin{figure}
\centering
\includegraphics{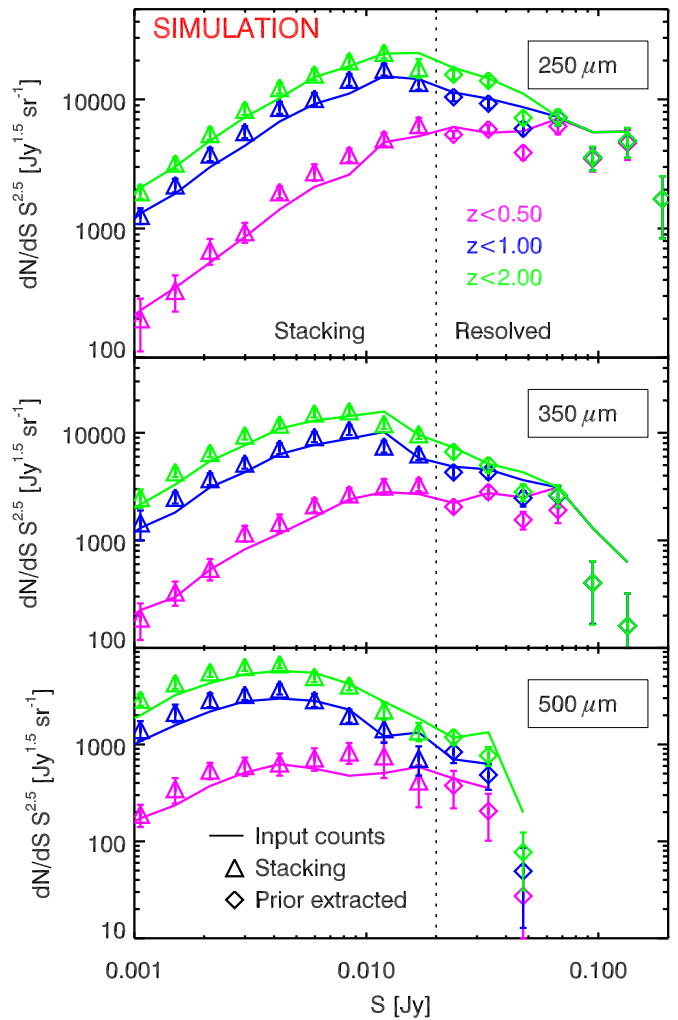}
\caption{\label{fig:simu} Validation of our method of measurement of the counts at 250\,$\mu$m (top), 350\,$\mu$m (center), and 500\,$\mu$m (bottom) from a simulation based on \citet{Bethermin2011} model. \textit{Solid lines:} input counts from the simulated catalog for various cuts in redshift. \textit{Diamonds}: resolved number counts measured using the same method as for the real data. \textit{Triangles}: number counts measured by stacking using the same method as for the real data.}
\end{figure}

The Fig.~\ref{fig:simu} shows the results of this simulation. The recovered counts (triangles and diamonds) nicely reproduce the shape of the counts. The flux density regime probed by stacking is well reproduced (reduced $\chi^2=1.4$ for 81 degrees of freedom). Paradoxically, the resolved counts are not as well reproduced, with some points deviating at more than 3-$\sigma$. At bright flux densities, our recovered counts are systematically lower than our results. It could be due to the fact that the extraction technique shares the flux of a bright source between several prior positions.\\

\section{Number counts}

\label{sect:number_counts}

\subsection{Results}

\begin{table*}
\centering
\begin{tabular}{lrrrrr}
\hline
\hline
\noalign{\vspace{2pt}}
Flux & \multicolumn{5}{c}{Normalized counts ($dN/dS \times S^{2.5}$)} \\
mJy & \multicolumn{5}{c}{Jy$^{1.5}$sr$^{-1}$} \\
 & All & $0<z<0.5$ & $0.5<z<1$ & $1<z<2$ & $z>2$ \\
 \noalign{\vspace{2pt}}
\hline
\noalign{\vspace{2pt}}
\multicolumn{6}{c}{Stacking (GOODS-N)}\\
\noalign{\vspace{2pt}}
\hline
\noalign{\vspace{2pt}}
2.1 & 4989 $\pm$ 824 & 699 $\pm$ 220 & 1828 $\pm$ 553 & 1392 $\pm$ 445 & 1068 $\pm$ 356 \\
3.0 & 7261 $\pm$ 1199 & 994 $\pm$ 250 & 2714 $\pm$ 941 & 2013 $\pm$ 542 & 1538 $\pm$ 442 \\
4.2 & 10405 $\pm$ 1337 & 1499 $\pm$ 307 & 3856 $\pm$ 1035 & 2812 $\pm$ 598 & 2237 $\pm$ 512 \\
\noalign{\vspace{2pt}}
\hline
\noalign{\vspace{2pt}}
\multicolumn{6}{c}{Stacking (COSMOS)}\\
\noalign{\vspace{2pt}}
\hline
\noalign{\vspace{2pt}}
6.0 & 22037 $\pm$ 3228 & 3086 $\pm$ 1082 & 6116 $\pm$ 1575 & 8918 $\pm$ 2273 & 3915 $\pm$ 1265 \\
8.4 & 25787 $\pm$ 2704 & 3759 $\pm$ 906 & 7463 $\pm$ 1389 & 9969 $\pm$ 1851 & 4594 $\pm$ 1064 \\
11.9 & 29044 $\pm$ 2168 & 4574 $\pm$ 728 & 8834 $\pm$ 1284 & 10727 $\pm$ 1345 & 4907 $\pm$ 843 \\
16.8 & 31574 $\pm$ 2527 & 5503 $\pm$ 723 & 9899 $\pm$ 1516 & 11185 $\pm$ 1579 & 4985 $\pm$ 1035 \\
\noalign{\vspace{2pt}}
\hline
\noalign{\vspace{2pt}}
\multicolumn{6}{c}{Resolved (COSMOS)}\\
\noalign{\vspace{2pt}}
\hline
\noalign{\vspace{2pt}}
23.8 & 23851 $\pm$ 1595 & 5558 $\pm$ 733 & 6694 $\pm$ 760 & 7516 $\pm$ 1076 & 4082 $\pm$ 519 \\
33.6 & 20926 $\pm$ 1654 & 5349 $\pm$ 804 & 4875 $\pm$ 690 & 7379 $\pm$ 1151 & 3323 $\pm$ 536 \\
47.4 & 12653 $\pm$ 1345 & 3878 $\pm$ 762 & 4036 $\pm$ 752 & 3114 $\pm$ 688 & 1623 $\pm$ 433 \\
67.0 & 8319 $\pm$ 1337 & 3813 $\pm$ 928 & 2281 $\pm$ 681 & 1777 $\pm$ 613 & 447 $\pm$ 289 \\
94.6 & 5780 $\pm$ 1431 & 3606 $\pm$ 1142 & 596 $\pm$ 432 & 1279 $\pm$ 681 & 298 $\pm$ 304 \\
133.7 & 528 $\pm$ 532 & - & 528 $\pm$ 532 & - & - \\
188.8 & 1306 $\pm$ 1084 & 1306 $\pm$ 1084 & - & - & - \\
\noalign{\vspace{2pt}}
\hline
\end{tabular}
\caption{Number counts at 250\,$\micro$m. The errors take into account the statistical uncertainties, including the clustering effect, and the uncertainties in the completeness corrections. For the points measured by stacking, we also take into account the uncertainties in the colors and the scatter. The uncertainties in the SPIRE absolute calibration are neglected here.}
\end{table*}

\begin{table*}
\centering
\begin{tabular}{lrrrrr}
\hline
\hline
\noalign{\vspace{2pt}}
Flux & \multicolumn{5}{c}{Normalized counts ($dN/dS \times S^{2.5}$)} \\
mJy & \multicolumn{5}{c}{Jy$^{1.5}$sr$^{-1}$} \\
 & All & $0<z<0.5$ & $0.5<z<1$ & $1<z<2$ & $z>2$ \\
 \noalign{\vspace{2pt}}
\hline
\noalign{\vspace{2pt}}
\multicolumn{6}{c}{Stacking (GOODS-N)}\\
\noalign{\vspace{2pt}}
\hline
\noalign{\vspace{2pt}}
2.1 & 4709 $\pm$ 1342 & 510 $\pm$ 159 & 1695 $\pm$ 572 & 1341 $\pm$ 450 & 1162 $\pm$ 1117 \\
3.0 & 6949 $\pm$ 1167 & 761 $\pm$ 234 & 2598 $\pm$ 870 & 1912 $\pm$ 481 & 1676 $\pm$ 564 \\
4.2 & 9964 $\pm$ 1396 & 1115 $\pm$ 363 & 3802 $\pm$ 1016 & 2587 $\pm$ 520 & 2459 $\pm$ 715 \\
\noalign{\vspace{2pt}}
\hline
\noalign{\vspace{2pt}}
\multicolumn{6}{c}{Stacking (COSMOS)}\\
\noalign{\vspace{2pt}}
\hline
\noalign{\vspace{2pt}}
6.0 & 21510 $\pm$ 3858 & 2143 $\pm$ 525 & 5083 $\pm$ 1309 & 8691 $\pm$ 2354 & 5592 $\pm$ 2711 \\
8.4 & 23820 $\pm$ 3174 & 2458 $\pm$ 431 & 5715 $\pm$ 996 & 9627 $\pm$ 1997 & 6020 $\pm$ 2216 \\
11.9 & 24402 $\pm$ 2274 & 2638 $\pm$ 462 & 6175 $\pm$ 1001 & 9875 $\pm$ 1445 & 5713 $\pm$ 1366 \\
16.8 & 24229 $\pm$ 3158 & 2579 $\pm$ 790 & 6115 $\pm$ 1648 & 9953 $\pm$ 2208 & 5581 $\pm$ 1325 \\
\noalign{\vspace{2pt}}
\hline
\noalign{\vspace{2pt}}
\multicolumn{6}{c}{Resolved (COSMOS)}\\
\noalign{\vspace{2pt}}
\hline
\noalign{\vspace{2pt}}
23.8 & 18652 $\pm$ 1605 & 1967 $\pm$ 434 & 3660 $\pm$ 643 & 8442 $\pm$ 1215 & 4581 $\pm$ 703 \\
33.6 & 15285 $\pm$ 1448 & 1927 $\pm$ 484 & 3600 $\pm$ 708 & 4995 $\pm$ 838 & 4760 $\pm$ 811 \\
47.4 & 9092 $\pm$ 1187 & 927 $\pm$ 346 & 1606 $\pm$ 467 & 3929 $\pm$ 828 & 2628 $\pm$ 620 \\
67.0 & 3487 $\pm$ 828 & 728 $\pm$ 390 & 327 $\pm$ 234 & 1527 $\pm$ 553 & 904 $\pm$ 416 \\
94.6 & 1163 $\pm$ 630 & 354 $\pm$ 351 & 98 $\pm$ 160 & 710 $\pm$ 497 & - \\
133.7 & 170 $\pm$ 273 & - & - & - & 170 $\pm$ 273 \\
\noalign{\vspace{2pt}}
\hline
\end{tabular}
\caption{Number counts at 350\,$\micro$m.}
\end{table*}

\begin{table*}
\centering
\begin{tabular}{lrrrrr}
\hline
\hline
\noalign{\vspace{2pt}}
Flux & \multicolumn{5}{c}{Normalized counts ($dN/dS \times S^{2.5}$)} \\
mJy & \multicolumn{5}{c}{Jy$^{1.5}$sr$^{-1}$} \\
 & All & $0<z<0.5$ & $0.5<z<1$ & $1<z<2$ & $z>2$ \\
 \noalign{\vspace{2pt}}
\hline
\noalign{\vspace{2pt}}
\multicolumn{6}{c}{Stacking (GOODS-N)}\\
\noalign{\vspace{2pt}}
\hline
\noalign{\vspace{2pt}}
2.1 & 3465 $\pm$ 864 & 368 $\pm$ 178 & 1262 $\pm$ 394 & 836 $\pm$ 217 & 998 $\pm$ 716 \\
3.0 & 5216 $\pm$ 783 & 493 $\pm$ 244 & 1966 $\pm$ 500 & 1220 $\pm$ 351 & 1535 $\pm$ 424 \\
4.2 & 7244 $\pm$ 1089 & 726 $\pm$ 360 & 2823 $\pm$ 631 & 1592 $\pm$ 529 & 2102 $\pm$ 615 \\
\noalign{\vspace{2pt}}
\hline
\noalign{\vspace{2pt}}
\multicolumn{6}{c}{Stacking (COSMOS)}\\
\noalign{\vspace{2pt}}
\hline
\noalign{\vspace{2pt}}
6.0 & 12170 $\pm$ 1764 & 750 $\pm$ 263 & 2381 $\pm$ 562 & 4444 $\pm$ 817 & 4594 $\pm$ 1435 \\
8.4 & 11446 $\pm$ 1716 & 596 $\pm$ 322 & 2107 $\pm$ 783 & 4481 $\pm$ 872 & 4260 $\pm$ 1210 \\
11.9 & 9917 $\pm$ 2089 & 465 $\pm$ 380 & 1586 $\pm$ 1000 & 3976 $\pm$ 1363 & 3888 $\pm$ 1167 \\
16.8 & 7540 $\pm$ 2665 & 358 $\pm$ 443 & 1055 $\pm$ 1192 & 3003 $\pm$ 1817 & 3122 $\pm$ 1478 \\
\noalign{\vspace{2pt}}
\hline
\noalign{\vspace{2pt}}
\multicolumn{6}{c}{Resolved (COSMOS)}\\
\noalign{\vspace{2pt}}
\hline
\noalign{\vspace{2pt}}
23.8 & 6298 $\pm$ 675 & 602 $\pm$ 220 & 1023 $\pm$ 278 & 2258 $\pm$ 343 & 2413 $\pm$ 460 \\
33.6 & 4548 $\pm$ 656 & 248 $\pm$ 138 & 483 $\pm$ 191 & 1493 $\pm$ 329 & 2322 $\pm$ 516 \\
47.4 & 1143 $\pm$ 343 & - & 130 $\pm$ 117 & 549 $\pm$ 238 & 463 $\pm$ 218 \\
67.0 & 343 $\pm$ 251 & - & - & 182 $\pm$ 185 & 160 $\pm$ 169 \\
94.6 & 202 $\pm$ 230 & - & - & 100 $\pm$ 162 & 101 $\pm$ 163 \\
\noalign{\vspace{2pt}}
\hline
\end{tabular}
\caption{Number counts at 500\,$\micro$m.}
\end{table*}

\label{sect:mes_nc}

\begin{figure*}
\centering
\includegraphics[width=16cm]{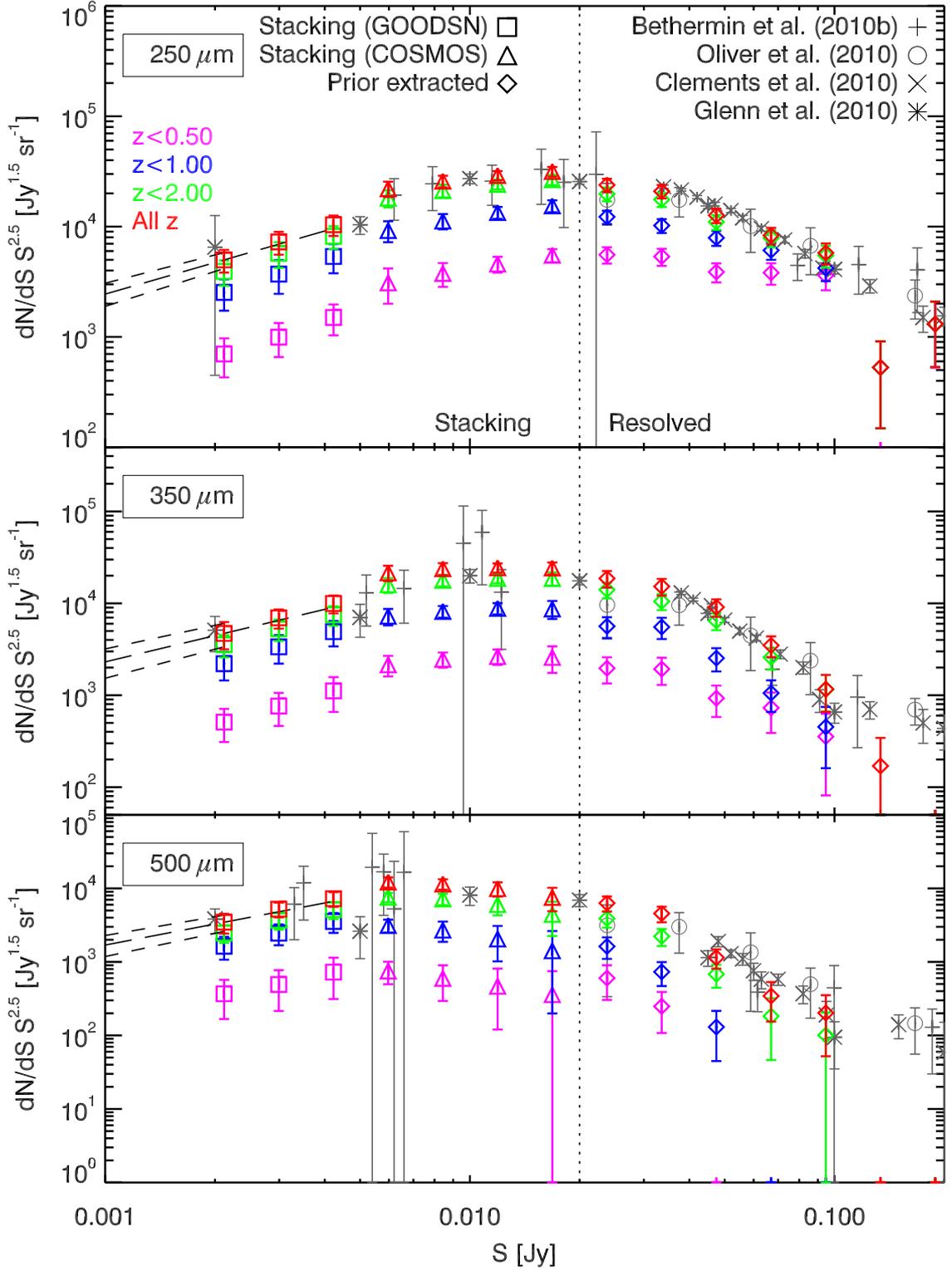}
\caption{\label{fig:counts} Number counts at 250\,$\micro$m (top panel), 350\,$\micro$m (middle panel), and 500\,$\micro$m (bottom panel). The contribution of $z<0.5$, $z<1$, $z<2$, and all sources are plotted in violet, blue, green and red, respectively. \textit{Squares}: points from stacking in GOODS-N. \textit{Triangles}: points from stacking in COSMOS. \textit{Diamonds}: points measured from source extraction using priors. \textit{Vertical dotted line}: 4-$\sigma$ confusion limit. \textit{Long and short dashed lines}: extrapolation of the counts and 1-$\sigma$ confidence region (see Sect.~\ref{sect:total_cib}). \textit{Plus symbols}: \citet{Bethermin2010b} measurements using BLAST data. \textit{Circles}: \citet{Oliver2010} measurements from resolved sources in the HerMES science demonstration phase data (\textit{Herschel}/SPIRE). \textit{Crosses}: \citet{Clements2010} measurements from resolved sources in the SPIRE H-ATLAS science demonstration phase data (\textit{Herschel}/SPIRE). \textit{Asterisks}: \citet{Glenn2010} measurements from $P(D)$ analysis of the HerMES science demonstration phase data (\textit{Herschel}/SPIRE).}
\end{figure*}

From the extraction with priors presented in Sect.~\ref{sect:prior_method}, we build number counts per redshift slice down to 20~mJy at all three SPIRE wavelengths. Thanks to the stacking of the 24\,$\micro$m sources in the COSMOS and GOODS-N fields, we reach 6~mJy and 2~mJy, respectively. We checked that the counts deduced from stacking analysis are in agreement with resolved counts above 20~mJy, but they have larger uncertainties than the resolved ones. The COSMOS and GOODS-N counts deduced by stacking analysis are also in agreement where they overlap, but with much smaller uncertainties in COSMOS due to the size of this field. We thus use GOODS-N points only at faint flux densities, which the COSMOS data does not constrain. Fig.~\ref{fig:counts} and \ref{fig:compare_counts} show our results. The points obtained by stacking in the GOODS-N and COSMOS fields disagree at 2$\sigma$ in the $1<z<2$ bin at all SPIRE wavelengths (see Fig.~\ref{fig:compare_counts}), which could be due to field-to-field variance.\\

The depth and small error bars of our counts enable us to detect with high significance the peak of the Euclidian normalized counts near 15, 10 and 5~mJy at 250\,$\micro$m, 350\,$\micro$m, and 500\,$\micro$m, respectively. This maximum was seen at 250\,$\micro$m and 350\,$\micro$m by \citet{Glenn2010}. With our new results, we are able to detect this maximum at 500~$\mu$m. We also start to see a power-law behavior below this peak, which was seen previously only up to 160\,$\micro$m \citep{Papovich2004,Bethermin2010a,Berta2011}. Nevertheless, the significance of this detection is hard to estimate because of the correlation between the points obtained by stacking.\\

\subsection{Comparison with the previous measurements}

We have compared our total counts with previous measurements (cf. Fig.~\ref{fig:counts}). At high flux densities (S$>$20~mJy), our counts agree with the counts measured from resolved sources of \citet{Bethermin2010b} in BLAST, \citet{Oliver2010} in SPIRE/HerMES SDP fields, and \citet{Clements2010} in the SPIRE/H-ATLAS SDP field. Our measurements are also in agreement with the stacking analysis of \citet{Bethermin2010b} of the BLAST data. Our new stacking analysis of the SPIRE data reduces the uncertainties by about a factor 5 compared for the BLAST data. Finally, we agree with the $P(D)$ analysis of \citet{Glenn2010}, except for the 6~mJy points at 250\,$\mu$m and 500\,$\micro$m which disagrees by about 2 $\sigma$ with our measurements. Due to the number of points compared (21), we expect to have about 2 points with 2 $\sigma$ difference, so this is not significant. 

The good agreement between the counts produced by the stacking and the $P(D)$ analysis confirms the accuracy of these two statistical methods. It also suggests that the galaxies seen in the mid-IR are a good tracer of the sources responsible for the sub-mm counts, and justifies a posteriori our choice to use the 24~$\mu$m sources as a prior. The mid-IR faint and far-IR bright population thus constitute a small contribution to the number counts.\\

\subsection{Comparison with the models}

\label{sect:nc_model}

\begin{figure*}
\centering
\includegraphics{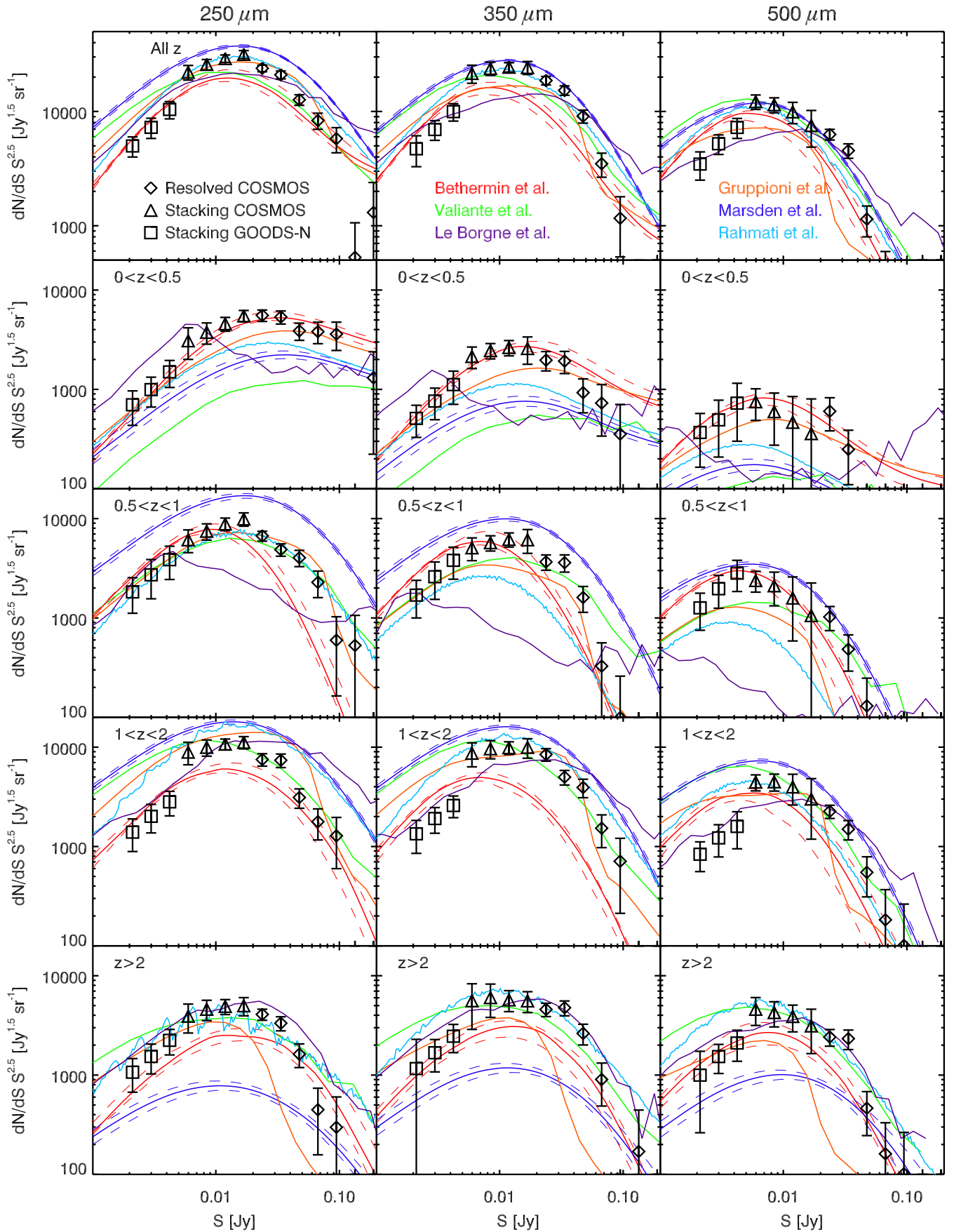}
\caption{\label{fig:compare_counts} Comparison between the observed number counts and the models at 250\,$\micro$m (left), 350\,$\micro$m (middle), and 500\,$\micro$m (right), for various redshift selections: all redshifts (top), $0<z<0.5$, $0.5<z<1$, $1<z<2$, and $z>2$ (bottom). \textit{Squares}: points from stacking in GOODS-N. \textit{Triangles}: points from stacking in COSMOS. \textit{Diamonds}: points measured from source extraction with priors. We have overplotted the models from \citet{Bethermin2011} in red, \citet{Valiante2009} in green, \citet{Le_Borgne2009} in violet, \citet{Gruppioni2011} in orange, \citet{Rahmati2011} in light blue and \citet{Marsden2010} in dark blue.}
\end{figure*}

In Fig.~\ref{fig:compare_counts}, we compare our results with a set of recent ($\ge$2009) evolutionary models:
\begin{itemize}

\item The \citet{Bethermin2011} model was presented in Sect.~\ref{sect:simu_desc}.

\item The \citet{Marsden2010} model is also a parametric model similar to the \citet{Bethermin2011} one, but using a different SED library, and taking into account the scatter in the temperature of the cold dust in the different galaxies.

\item \citet{Le_Borgne2009} carried out a non-parametric inversion of the counts assuming a single population \citep{Chary2001} to determine the evolution of the luminosity function with redshift.

\item The \citet{Valiante2009} model used a large library of starburst and AGNs templates. This model takes into account the scatter in the temperature of the sources. The parameters of the model were tuned manually.

\item The \citet{Gruppioni2011} model uses 5 separately evolving populations, including 3 populations of AGN.

\item The \citet{Rahmati2011} model is based on a modified \citet{Dale2002} library. This model takes into account the scatter in the temperature of the sources. It was fit to the 850\,$\micro$m counts and  redshift distribution.

\end{itemize}

Note that the \citet{Bethermin2011}, \citet{Gruppioni2011} and \citet{Rahmati2011} models were already tuned using recent \textit{Herschel} data, including the GOODS-N observations used here.\\

None of these models manages to fully reproduce our measurements. The \citet{Bethermin2011}, \citet{Gruppioni2011} and \citet{Rahmati2011} models are close to the data, and broadly reproduce the shape of the counts, but still deviate from the measurements by 3-$\sigma$. The \citet{Le_Borgne2009} and \citet{Valiante2009} models underestimate the contribution of $z<1$ sources to the counts. The \citet{Marsden2010} model overestimates the counts at high $z$ ($z>1$). Not surprisingly, models which use the most  recent \textit{Herschel} data and the redshift-dependent observables (redshift distributions, luminosity functions, etc.) provide the best match to our findings.\\

\section{Redshift distributions}

\label{sect:redshift_distributions}

\subsection{Results}

\label{sect:nz_res}

\begin{figure}
\centering
\includegraphics[width=9cm]{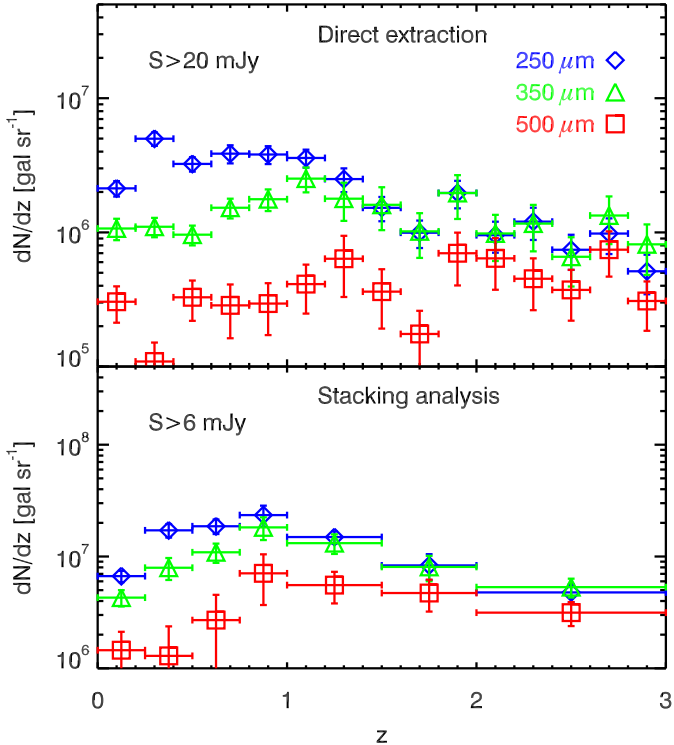}
\caption{\label{fig:Nz} Redshift distribution of S$_{\rm SPIRE}>20$~mJy (upper panel) and S$_{\rm SPIRE}>6$~mJy (lower panel) sources at 250\,$\micro$m (blue), 350\,$\micro$m (green), and 500\,$\micro$m (red) in the COSMOS field.}
\end{figure}

From the brighter sources extracted using the 24\,$\micro$m prior (see Sect.~\ref{sect:prior_method}), we have built the redshift distribution of the sources brighter than 20~mJy at 250\,$\micro$m, 350\,$\micro$m, and 500\,$\micro$m (see Fig.~\ref{fig:Nz} and Table~\ref{tab:Nz}). We find that the distribution of the resolved 250\,$\micro$m sources is almost flat up to $z \sim1$ and decreases significantly at higher redshift. At 350\,$\micro$m, the distribution peaks near $z\sim1$, and the distribution is flatter at high redshift. At 500~$\mu$m, the contribution of $z<1.5$ sources is smaller than at shorter wavelengths. Between $z=1.5$ and $z=3$, our measurements are compatible with a flat distribution, however the uncertainties are very large.\\

\begin{figure}
\centering
\includegraphics{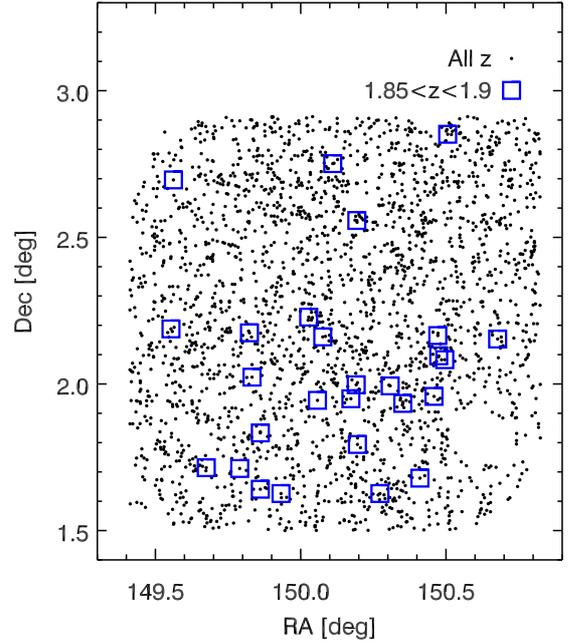}
\caption{\label{fig:overdensity} Spatial distribution of S$_{250}>$20~mJy sources in COSMOS. \textit{Black dots}: all redshifts. \textit{Blue boxes}: only sources in the 1.85$<z<$1.9 range.}
\end{figure}

At 250\,$\micro$m and 350~$\mu$m, we clearly see an excess in the $0.2<z<0.4$ and $1.8<z<2.0$ bins. The structure at $z=0.3$ in COSMOS is well known \citep{Scoville2007}. The excess near $z=1.9$ could also be explained by a large-scale structure. Fig.~\ref{fig:overdensity} shows the position of the sources in a thin redshift slice between $z=1.85$ and $z=1.9$. The sources are strongly concentrated in a 0.7$^\circ \times$0.7$^\circ$ region, corresponding to a physical size of about 20\,Mpc. It could be linked with the three candidate clusters of galaxies at $z\sim1.8$ found by \citet{Chiaberge2010} in the same field. Nevertheless, this overdensity could also be an artifact of the photometric redshifts. An effect of the polycyclic aromatic hydrocarbon (PAH) features redshifting into the 24\,$\micro$m band is possible althought less likely, because this should affect neighboring redshift bins, due to the band width of \textit{Spitzer} at 24\,$\micro$m ($\lambda / \Delta \lambda \sim 3$).\\

We also used the stacking analysis presented in Sect.~\ref{sect:spire_reconstruction} to estimate the redshift distribution of the S$_{\rm SPIRE}>$6\,mJy sources in COSMOS. We find a smaller relative contribution of $z<1$ sources than for the 20~mJy flux density cut at 250\,$\mu$m and 350\,$\mu$m. The behavior at $z>1$ is similar to that found for the 20\,mJy flux density cut.\\

\begin{table}
\centering
\begin{tabular}{lrrr}
\hline
\hline
\noalign{\vspace{2pt}}
redshift range& \multicolumn{3}{c}{dN/dz (in 10$^4\times$~gal.sr$^{-1}$)} \\
\noalign{\vspace{2pt}}
\hline
\noalign{\vspace{2pt}}
 & $S_{250}>20$~mJy & $S_{350}>20$~mJy & $S_{500}>20$~mJy \\
 \noalign{\vspace{2pt}}
\hline
\noalign{\vspace{2pt}}
$0.0<z<0.2$ & 212$\pm$28 & 106$\pm$19 & 30$\pm$9 \\
$0.2<z<0.4$ & 498$\pm$57 & 110$\pm$18 & 10$\pm$4 \\
$0.4<z<0.6$ & 323$\pm$39 & 96$\pm$16 & 32$\pm$10 \\
$0.6<z<0.8$ & 386$\pm$59 & 152$\pm$25 & 28$\pm$12 \\
$0.8<z<1.0$ & 380$\pm$58 & 176$\pm$32 & 29$\pm$12 \\
$1.0<z<1.2$ & 358$\pm$54 & 251$\pm$52 & 41$\pm$16 \\
$1.2<z<1.4$ & 249$\pm$50 & 178$\pm$56 & 63$\pm$30 \\
$1.4<z<1.6$ & 152$\pm$31 & 160$\pm$56 & 36$\pm$16 \\
$1.6<z<1.8$ & 99$\pm$23 & 101$\pm$37 & 17$\pm$8 \\
$1.8<z<2.0$ & 196$\pm$45 & 196$\pm$70 & 69$\pm$29 \\
$2.0<z<2.2$ & 95$\pm$24 & 98$\pm$37 & 63$\pm$26 \\
$2.2<z<2.4$ & 120$\pm$32 & 116$\pm$44 & 45$\pm$18 \\
$2.4<z<2.6$ & 74$\pm$21 & 65$\pm$26 & 37$\pm$15 \\
$2.6<z<2.8$ & 98$\pm$29 & 133$\pm$51 & 74$\pm$27 \\
$2.8<z<3.0$ & 51$\pm$16 & 81$\pm$33 & 30$\pm$12 \\
\noalign{\vspace{2pt}}
\hline
\noalign{\vspace{2pt}}
 & $S_{250}>6$~mJy & $S_{350}>6$~mJy & $S_{500}>6$~mJy \\
 \noalign{\vspace{2pt}}
\hline
\noalign{\vspace{2pt}}
$0.0<z<0.2$ & 667$\pm$76 & 428$\pm$71 & 145$\pm$66 \\
$0.2<z<0.5$ & 1713$\pm$236 & 793$\pm$176 & 129$\pm$107 \\
$0.5<z<0.8$ & 1865$\pm$277 & 1090$\pm$213 & 269$\pm$184 \\
$0.8<z<1.0$ & 2346$\pm$500 & 1821$\pm$417 & 707$\pm$339 \\
$1.0<z<1.5$ & 1493$\pm$183 & 1316$\pm$262 & 555$\pm$174 \\
$1.5<z<2.0$ & 832$\pm$216 & 807$\pm$206 & 471$\pm$149 \\
$2.0<z<3.0$ & 477$\pm$88 & 531$\pm$101 & 314$\pm$76 \\
$z>3.0$ & 11$\pm$1 & 11$\pm$1 & 9$\pm$1 \\
\noalign{\vspace{2pt}}
\hline
\end{tabular}
\caption{\label{tab:Nz} Redshift distribution of the SPIRE sources in COSMOS for various flux density cuts at the three SPIRE wavelengths.}
\end{table}

\subsection{Comparison with other measurements}

\citet{Chapin2010} studied the redshift distribution of isolated BLAST sources. Their redshift distributions cannot be normalized by the surface area (because of the isolation ctriterion), and the flux density cuts are different; nevertheless, the trends of their distributions and their evolution from 250\,$\micro$m to 500\,$\micro$m agrees with our findings.\\

\citet{Amblard2010} also produced a redshift distribution of $S_{350}>35$\,mJy sources in H-ATLAS from a \textit{Herschel} color-color diagram and using assumptions about the FIR/sub-mm SED of the sources. They found a strong peak at $z=2$, in complete disagreement with our distribution. This could be due to the fact that they required a 3 $\sigma$ detection at 250\,$\micro$m and 500\,$\micro$m, which correspond to a $S_{250}>21$\,mJy and $S_{500}>27$\,mJy. The 3-$\sigma$ criterion at 500\,$\micro$m tends to select high-redshift sources, because of the shape of the SEDs, the flux of low-$z$ sources decreases rapidly between 250\,$\mu$m and 500\,$\mu$m. The method is also strongly dependent on the dust temperatures of the sources assumed in their analysis, due to the degeneracy between dust temperature and redshift for thermal sources.\\

\subsection{Comparison with the models}

\begin{figure*}
\centering
\includegraphics{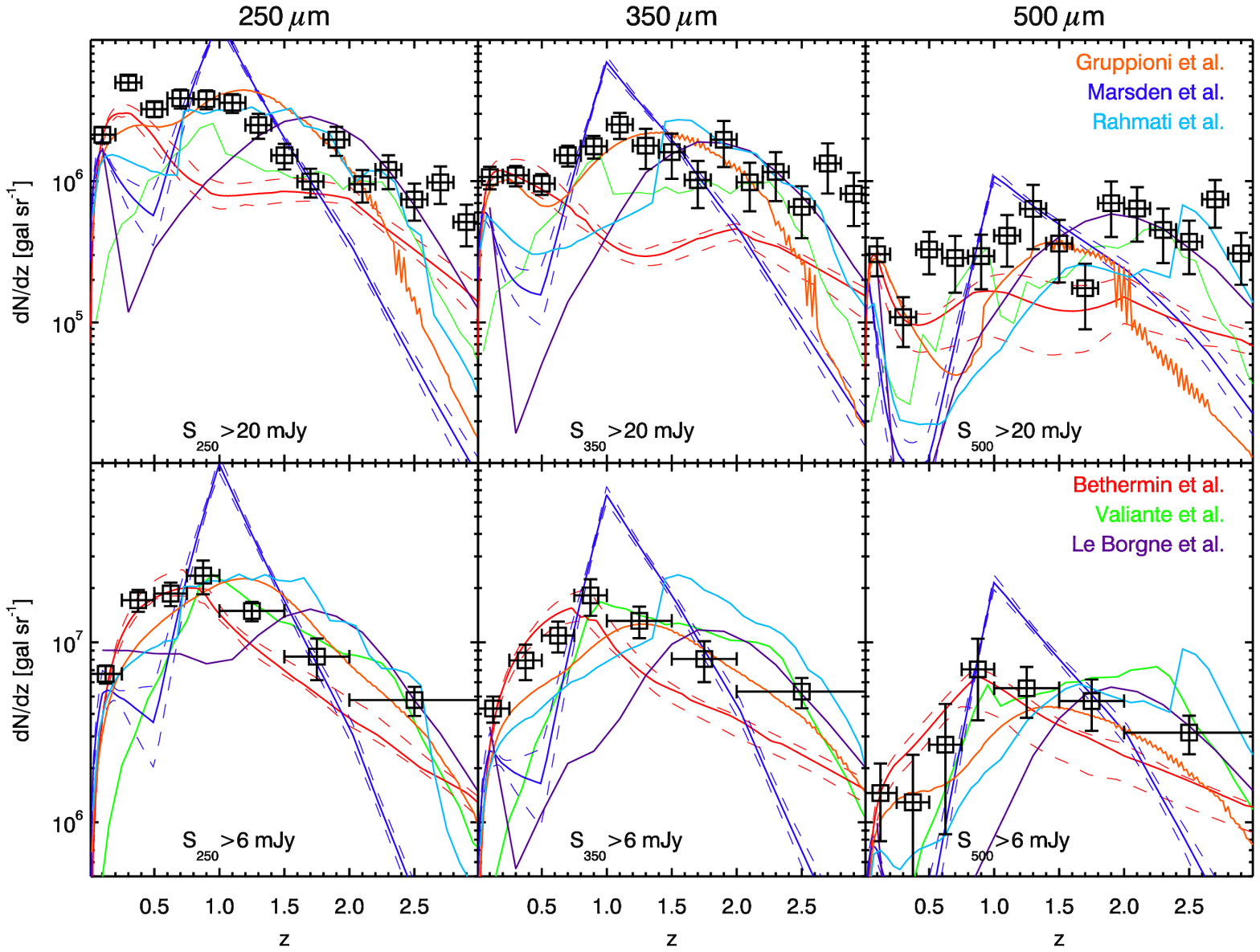}
\caption{\label{fig:compare_nz} Redshift distribution of the $S_{\rm SPIRE}>20$\,mJy (upper panels) and S$_{SPIRE}>6$~mJy (lower panels) sources at 250\,$\micro$m (left), 350\,$\micro$m (center), and 500\,$\micro$m (right). We overplot the models of \citet{Bethermin2011} in red, \citet{Valiante2009} in green, \citet{Le_Borgne2009} in violet, \citet{Gruppioni2011} in orange, \citet{Rahmati2011} in light blue and \citet{Marsden2010} in dark blue.}
\end{figure*}

We compared the measured redshift distributions with the predictions of the same models as in Sect.~\ref{sect:nc_model} (Fig.~\ref{fig:compare_nz}). Again no model manages to reproduce accurately the redshift distributions of the bright resolved sources ($S>20$\,mJy). Note however that \citet{Gruppioni2011} model reasonably fits the data at 250~$\mu$m and 350~$\mu$m at $z<2.5$. All the models without strong lensing predict a strong break in the redshift distributions at z$\sim$2.5, which is not present in the data. The \citet{Bethermin2011} model, which includes strong lensing, predicts a more consistent slope, although the normalization is not correct. It could be interpreted as a clue that the high redshift tail is due to lensed galaxies \citep[see e.g.][]{Vieira2010,Negrello2010}, but the contribution of lensed galaxies in the \citet{Bethermin2011} model is negligible for a flux density cut of 20~mJy\footnote{Note however that the lensed objects dominates the redshift distribution at $z>2$ for a flux density cut of 100~mJy.}.The redshift distribution of the faint sources ($S>6$\,mJy) are globally better modeled, a broad agreement being found with the \citet{Bethermin2011} and \citet{Gruppioni2011} models, which are fitted using the most recent data. The strong disagreement with the models at bright flux densities suggests that the bright end of the luminosity function and/or the SEDs of the brightest objects are not well modeled by the current studies. Our measurements therefore provide significant new constraints for such models.\\

\section{Cosmic infrared background}

\label{sect:cib}

\subsection{Contribution of the 24\,$\micro$m-selected sources to the CIB}

\begin{figure}
\centering
\includegraphics[width=8cm]{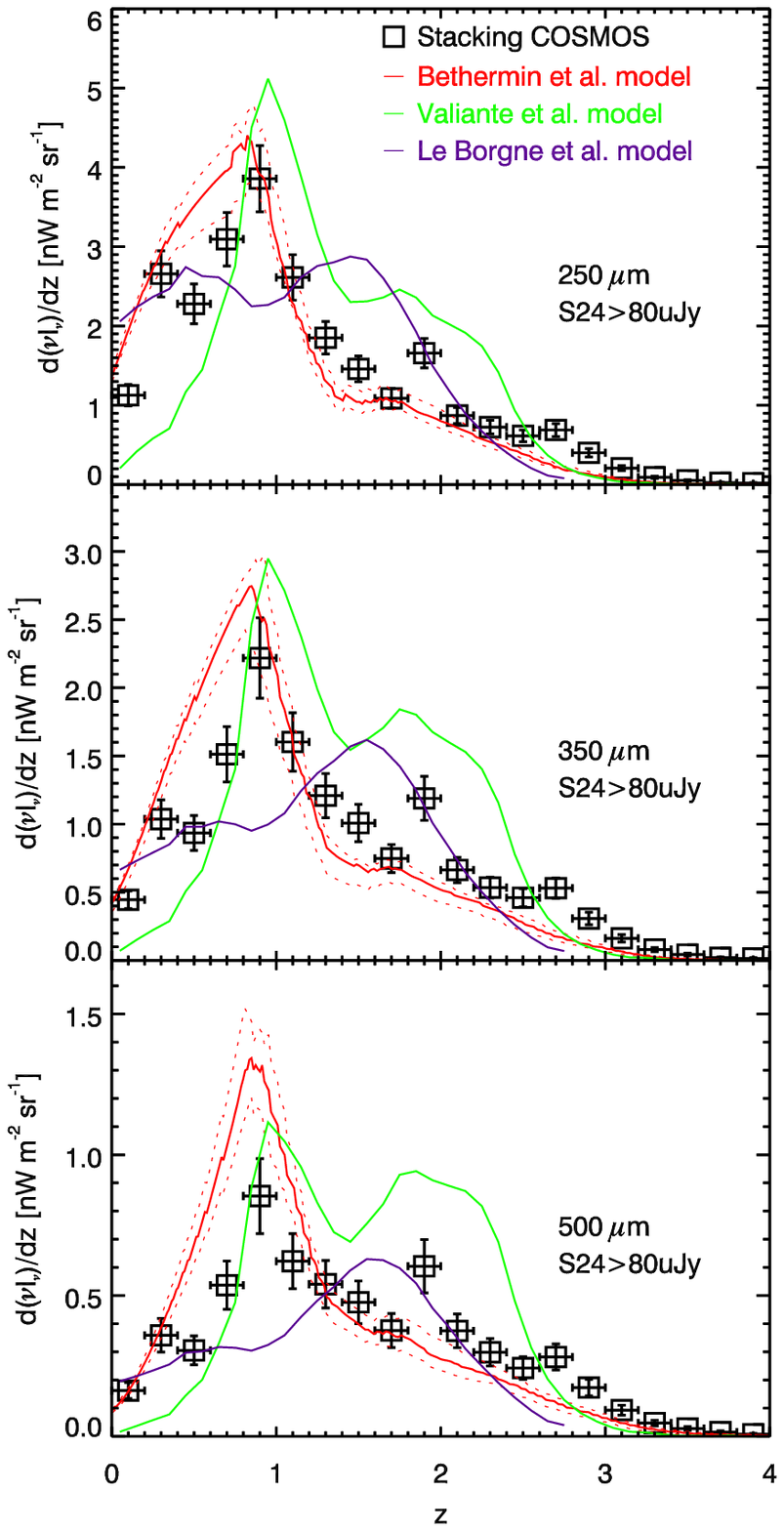}
\caption{\label{fig:dBdz24} Contribution of $S_{24}>80\,\mu$Jy sources to the CIB as a function of redshift at 250\,$\micro$m (top), 350\,$\micro$m (center), and 500\,$\micro$m (bottom). We overplotted the predictions of the \citet{Bethermin2011} (red), \citet{Valiante2009} (green), and \citet{Le_Borgne2009} (violet) models.}
\end{figure}

The differential contribution of the 24\,$\micro$m-selected sources to the CIB at longer wavelengths as a function of redshift is a relatively unbiased measurement and places tight constraints on evolution models. Measurements were performed at 70 and 160\,$\micro$m in \textit{Spitzer} data by \citet{Jauzac2011}, and at 250\,$\micro$m, 350\,$\micro$m, and 500\,$\micro$m by Viera et al. (in prep.). The latter were performed in GOODS-N, on a small area. We performed the same analysis in COSMOS, obtaining smaller uncertainties and a better resolution in redshift. To compute this overall observable, we have estimated the total surface brightness due to the $S_{24}>80\,\mu$Jy sources in redshift slices. The results of this stacking analysis is shown in Fig.~\ref{fig:dBdz24} and given in Table~\ref{tab:dBdz}.\\

As in Vieira et al. (in prep.), we find a peak near $z=1$. The relative contribution of the $z<1$ sources decreases with wavelength, and the contribution of $z>1$ sources increases. We observed 2 peaks at $z\sim0.3$ and $z\sim1.9$, probably associated with the overdensities discussed in Sect.~\ref{sect:nz_res}. We compared our results with the predictions of the three models which can take into account the 24\,$\micro$m selection among the six previously-compared ones. The \citet{Bethermin2011} model broadly reproduces our measurements. Nevertheless, it overpredicts by 3-$\sigma$ the observed values below $z=1$ at 500\,$\micro$m . The \citet{Valiante2009} model predicts a large bump near $z=2$, which is not seen. A smaller bump is predicted by the \citet{Le_Borgne2009} model. However, this model tends to underestimate the contribution of $z<1$ sources and overestimate the contribution of $z>1$ ones. The large bumps in the CIB contribution predicted by the \citet{Valiante2009} and \citet{Le_Borgne2009} models around $z \sim2$, caused by PAH features, are not seen in our data, although there is a single elevated point at $z\sim1.8$ in all bands. We are unsure as to the cause of this observed feature, but note that, given the width of the MIPS 24\,$\mu$m filter, we would expect any significant PAH contribution to affect multiple redshift bins instead of a single point.\\

\begin{table}
\centering
\begin{tabular}{lrrr}
\hline
\hline
\noalign{\vspace{2pt}}
redshift range & \multicolumn{3}{c}{$d(\nu I_\nu)/dz$ (in nW\,m$^{-2}$\,sr$^{-1}$)} \\
 & 250\,$\micro$m & 350\,$\micro$m & 500\,$\micro$m \\
 \noalign{\vspace{2pt}}
\hline
\noalign{\vspace{2pt}}
$0.0<z<0.2$ & 1.127$\pm$0.137 & 0.446$\pm$0.057 & 0.163$\pm$0.024 \\
$0.2<z<0.4$ & 2.657$\pm$0.295 & 1.037$\pm$0.123 & 0.359$\pm$0.048 \\
$0.4<z<0.6$ & 2.279$\pm$0.254 & 0.935$\pm$0.111 & 0.305$\pm$0.042 \\
$0.6<z<0.8$ & 3.094$\pm$0.339 & 1.512$\pm$0.175 & 0.537$\pm$0.068 \\
$0.8<z<1.0$ & 3.857$\pm$0.421 & 2.218$\pm$0.253 & 0.853$\pm$0.104 \\
$1.0<z<1.2$ & 2.612$\pm$0.288 & 1.602$\pm$0.185 & 0.622$\pm$0.077 \\
$1.2<z<1.4$ & 1.851$\pm$0.206 & 1.208$\pm$0.140 & 0.540$\pm$0.066 \\
$1.4<z<1.6$ & 1.459$\pm$0.165 & 1.008$\pm$0.118 & 0.476$\pm$0.060 \\
$1.6<z<1.8$ & 1.088$\pm$0.125 & 0.748$\pm$0.090 & 0.376$\pm$0.048 \\
$1.8<z<2.0$ & 1.658$\pm$0.187 & 1.189$\pm$0.139 & 0.604$\pm$0.074 \\
$2.0<z<2.2$ & 0.871$\pm$0.102 & 0.664$\pm$0.080 & 0.375$\pm$0.048 \\
$2.2<z<2.4$ & 0.725$\pm$0.086 & 0.534$\pm$0.066 & 0.299$\pm$0.039 \\
$2.4<z<2.6$ & 0.615$\pm$0.075 & 0.460$\pm$0.058 & 0.242$\pm$0.033 \\
$2.6<z<2.8$ & 0.689$\pm$0.082 & 0.531$\pm$0.066 & 0.282$\pm$0.037 \\
$2.8<z<3.0$ & 0.401$\pm$0.051 & 0.308$\pm$0.041 & 0.173$\pm$0.024 \\
$3.0<z<3.2$ & 0.207$\pm$0.031 & 0.162$\pm$0.024 & 0.093$\pm$0.015 \\
$3.2<z<3.4$ & 0.092$\pm$0.016 & 0.080$\pm$0.014 & 0.047$\pm$0.009 \\
$3.4<z<3.6$ & 0.052$\pm$0.011 & 0.043$\pm$0.009 & 0.028$\pm$0.006 \\
$3.6<z<3.8$ & 0.023$\pm$0.007 & 0.020$\pm$0.006 & 0.015$\pm$0.005 \\
$3.8<z<4.0$ & 0.018$\pm$0.006 & 0.015$\pm$0.005 & 0.007$\pm$0.003 \\
\noalign{\vspace{2pt}}
\hline
\end{tabular}
\caption{\label{tab:dBdz} Differential contribution of S$_{24}>80 \mu$Jy sources to the CIB as a function of redshift.}
\end{table}

\subsection{Properties of the CIB }

The contribution of the 24\,$\micro$m sources to the CIB is an interesting quantity for models. Nevertheless, we want to have constraints on the total contribution of the galaxies to the CIB, even if the uncertainties are larger. These constraints can be derived by integrating and extrapolating our new SPIRE number counts.\\

\subsubsection{Estimate of the contribution to the CIB from galaxies}

\label{sect:total_cib}

We integrated our counts for different cuts in flux density density, assuming the data points are connected by power-laws. The contribution to the CIB of the sources brighter than the brightest constrained flux bin is less than 2\% \citep{Bethermin2010b}, and is neglected. We estimated our error bars using a Monte Carlo method. We used the distribution of recovered values of the CIB to compute the confidence interval. We adopted this method down to faintest flux density probed by stacking. In order to take into account cosmic variance, we combined the statistical uncertainties with the 15\% level of the large scale fluctuations measured by \citet{Planck_CIB}.\\

We also extrapolated the contribution of the sources fainter than the limit of our counts. The typical faint-end slope of the infrared counts\footnote{The slope $\alpha$ of the counts is defined by $dN/dS \propto S^\alpha$.} lies in a range between -1.45 and -1.65 \citep{Papovich2004,Bethermin2010a,Berta2011}. This is also the case for our input redshift catalog, even if we select only a redshift slice. We thus assumed a slope of -1.55$\pm$0.10 to estimate the contribution of the flux density fainter than the limit of the stacking analysis. The errors are estimated using a MC process, which takes into account the uncertainties in the faint-end slope. By integrating our number counts extrapolated down to zero flux density, we find a total contribution of the galaxies to the CIB of 10.13$_{-2.33}^{+2.60}$\,nW\,m$^{-2}$\,sr$^{-1}$, 6.46$_{-1.57}^{+1.74}$\,nW\,m$^{-2}$\,sr$^{-1}$, and 2.80$_{-0.81}^{+0.93}$\,nW\,m$^{-2}$\,sr$^{-1}$ at 250\,$\micro$m, 350\,$\micro$m, and 500\,$\micro$m, respectively. These values agree at better than 1\,$\sigma$ with the FIRAS absolute measurements performed by \citet{Fixsen1998} and \citet{Lagache2000}.\\

We estimated the fraction of the CIB resolved into individual sources ($S>20$\,mJy) using our estimation of the total CIB coming from our extrapolation of the number counts down to zero flux density. We found 15\%, 11\% and 5\% at 250\,$\micro$m, 350\,$\micro$m, and 500\,$\micro$m, respectively. When we go down to 2~mJy (the limit of the stacking analysis), we resolve 73\%, 69\%, and 55\% of the CIB, respectively.

\begin{figure*}
\centering
\includegraphics[width=18cm]{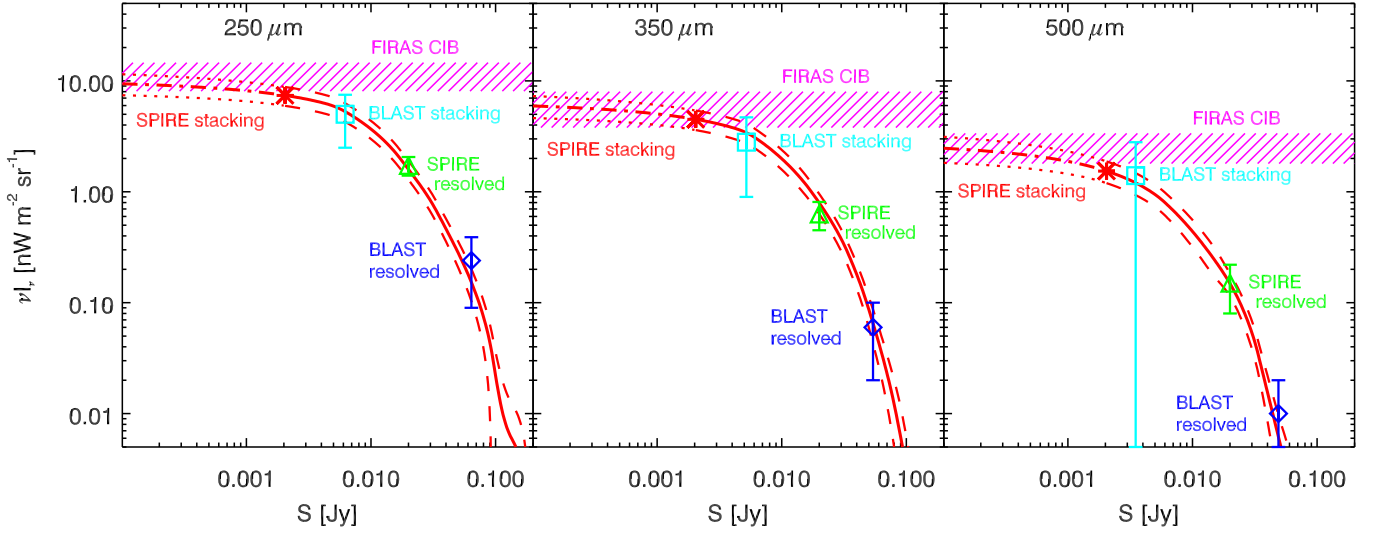}
\caption{\label{fig:cib_int} Cumulative contribution to the CIB as a function of the flux density cut at 250\,$\micro$m (left), 350\,$\micro$m (center), and 500\,$\micro$m (right). \textit{Red}: cumulative contribution from our counts. The asterisks represents the fraction resolved at the limit used for our analysis. \textit{Cyan}: contribution of the BLAST sources probed by stacking \citep{Bethermin2010b}. \textit{Green}: contribution of the sources resolved by SPIRE \citep{Oliver2010}. \textit{Blue}: contribution of the sources resolved by BLAST \citep{Bethermin2010b}. \textit{Violet hatched region}: FIRAS absolute measurement of the CIB; a region is hatched here if it is in the 1-$\sigma$ confidence region of \citet{Fixsen1998} or \citet{Lagache2000}.}
\end{figure*}

Fig.~\ref{fig:cib_int} shows the cumulative contribution to the CIB as a function of the flux density cut. We have compared our results with the fraction resolved by previous shallower analyses \citep{Bethermin2010b,Oliver2010}, and find a 1\,$\sigma$ agreement.\\ 

\begin{table*}
\centering
\begin{tabular}{lrrrrrrr}
\hline
\hline
\noalign{\vspace{2pt}}
& \multicolumn{2}{c}{Resolved ($S>20$\,mJy)} & \multicolumn{2}{c}{Stacking ($S>2$\,mJy)} & \multicolumn{1}{c}{Total extrapolated} & \multicolumn{2}{c}{Absolute measurements} \\
\noalign{\vspace{2pt}}
\hline
\noalign{\vspace{2pt}}
Wavelength & level & fraction & level & fraction & level & \citet{Fixsen1998} & \citet{Lagache2000}\\ 
$\micro$m & nW\,m$^{-2}$\,sr$^{-1}$ & & nW\,m$^{-2}$\,sr$^{-1}$ & & nW\,m$^{-2}$\,sr$^{-1}$ & nW\,m$^{-2}$\,sr$^{-1}$ & nW\,m$^{-2}$\,sr$^{-1}$\\
\noalign{\vspace{2pt}}
\hline
\noalign{\vspace{2pt}}
250 & 1.55$_{-0.30}^{+0.30}$ & 15\% & 7.40$_{-1.43}^{+1.42}$ & 73\% & 10.13$_{-2.33}^{+2.60}$ & 10.40$\pm$2.30 & 11.75$\pm$2.90 \\
\noalign{\vspace{2pt}}
350 & 0.77$_{-0.16}^{+0.16}$ & 11\% & 4.50$_{-0.90}^{+0.90}$ & 69\% & 6.46$_{-1.57}^{+1.74}$ & 5.40$\pm$1.60 & 6.43$\pm$1.59 \\
\noalign{\vspace{2pt}}
500 & 0.14$_{-0.03}^{+0.03}$ & 5\% & 1.54$_{-0.34}^{+0.34}$ & 55\% & 2.80$_{-0.81}^{+0.93}$ & 2.40$\pm$0.60 & 2.70$\pm$0.67 \\
\noalign{\vspace{2pt}}
\hline
\end{tabular}
\caption{\label{tab:cibsups} Summary of the contribution to the CIB for various flux density cuts, and comparison with the absolute measurements (which themselves have large uncertainties).}
\end{table*}

\subsubsection{CIB build-up as a function of redshift}

\begin{figure}
\centering
\includegraphics[width=8cm]{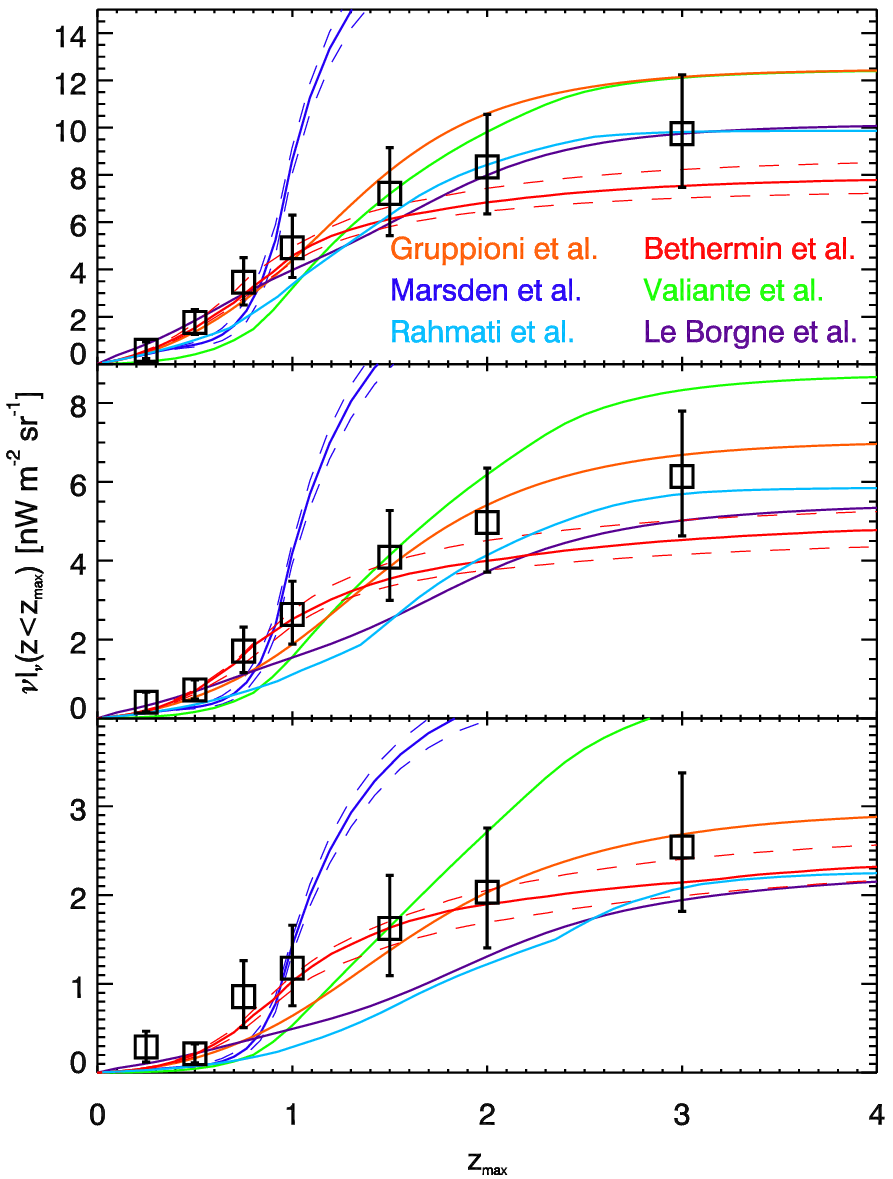}
\caption{\label{fig:dBdztot} Cumulative contribution to the CIB as a function of redshift at 250\,$\micro$m (top), 350\,$\micro$m (center), and 500\,$\micro$m (bottom), and comparison with the models of \citet{Bethermin2011} in red, \citet{Valiante2009} in green, \citet{Le_Borgne2009} in violet, \citet{Gruppioni2011} in orange, \citet{Rahmati2011} in light blue and \citet{Marsden2010} in dark blue.}
\end{figure}

From our cumulative number counts as a function of redshift (see Sect.~\ref{sect:mes_nc}), we can extrapolate the CIB emitted below a given redshift, following the methods presented in Sect.~\ref{sect:total_cib}. The results are presented in Fig.~\ref{fig:dBdztot} and Table \ref{tab:CIB_buildup}. The redshift at which half of the CIB is emitted is 1.04, 1.20, and 1.25, at 250\,$\micro$m, 350\,$\micro$m, and 500\,$\micro$m, respectively. For comparison, \citet{Le_Floch2009} measured value $z=1.08$ at 24\,$\micro$m. \citet{Berta2011} performed the same type of measurement, but considering only the resolved CIB, and found $z=0.58$, $z=0.67$, and $z=0.73$, at 70\,$\micro$m, 100\,$\micro$m, and 160\,$\micro$m, respectively. As expected, the CIB at longer wavelengths is emitted at higher redshift. The predictions of different models are also shown. The \citet{Marsden2010} and \citet{Valiante2009} models strongly overpredict the contribution of $z>1$ sources. The \citet{Le_Borgne2009} and \citet{Rahmati2011} models slightly underpredict the contribution of $z<2$ sources at 350~$\mu$m and 500~$\mu$m. The \citet{Gruppioni2011} model agrees at 1$\sigma$ with the measurements, except a 1.5$\sigma$ underprediction at z$\sim$1 at 500~$\mu$m. The \citet{Bethermin2011} models agrees at 1$\sigma$ with this measurement. Note, however, that it underestimates by 1$\sigma$ the contribution of $z<2$ sources to the CIB at 250~$\mu$m and 350~$\mu$m.

\begin{table}
\centering
\begin{tabular}{lrrr}
\hline
\hline
\noalign{\vspace{2pt}}
$z_{\rm max}$ & \multicolumn{3}{c}{$\nu I_\nu(z<z_{\rm max}$) (in nW\,m$^{-2}$\,sr$^{-1}$)} \\
 & 250\,$\micro$m & 350\,$\micro$m & 500\,$\micro$m \\
 \noalign{\vspace{2pt}}
\hline
\noalign{\vspace{2pt}}
0.2 & 0.6$_{-0.3}^{+0.4}$ & 0.4$_{-0.2}^{+0.3}$ & 0.3$_{-0.2}^{+0.2}$ \\
\noalign{\vspace{2pt}}
0.5 & 1.8$_{-0.5}^{+0.5}$ & 0.7$_{-0.2}^{+0.3}$ & 0.2$_{-0.1}^{+0.1}$ \\
\noalign{\vspace{2pt}}
0.8 & 3.5$_{-1.0}^{+1.0}$ & 1.7$_{-0.5}^{+0.6}$ & 0.9$_{-0.3}^{+0.4}$ \\
\noalign{\vspace{2pt}}
1.0 & 4.9$_{-1.3}^{+1.4}$ & 2.6$_{-0.8}^{+0.8}$ & 1.2$_{-0.4}^{+0.5}$ \\
\noalign{\vspace{2pt}}
1.5 & 7.2$_{-1.8}^{+1.9}$ & 4.1$_{-1.1}^{+1.2}$ & 1.6$_{-0.5}^{+0.6}$ \\
\noalign{\vspace{2pt}}
2.0 & 8.3$_{-2.0}^{+2.2}$ & 5.0$_{-1.3}^{+1.4}$ & 2.0$_{-0.6}^{+0.7}$ \\
\noalign{\vspace{2pt}}
3.0 & 9.7$_{-2.3}^{+2.5}$ & 6.1$_{-1.5}^{+1.7}$ & 2.5$_{-0.7}^{+0.8}$ \\
\noalign{\vspace{2pt}}
\hline
\end{tabular}
\caption{\label{tab:CIB_buildup} CIB build-up as a function of redshift at 250\,$\micro$m, 350\,$\micro$m, and 500\,$\micro$m.}
\end{table}

\subsection{Spectral energy distribution of the CIB and total integrated CIB}

\begin{figure*}
\centering
\includegraphics[width=19cm]{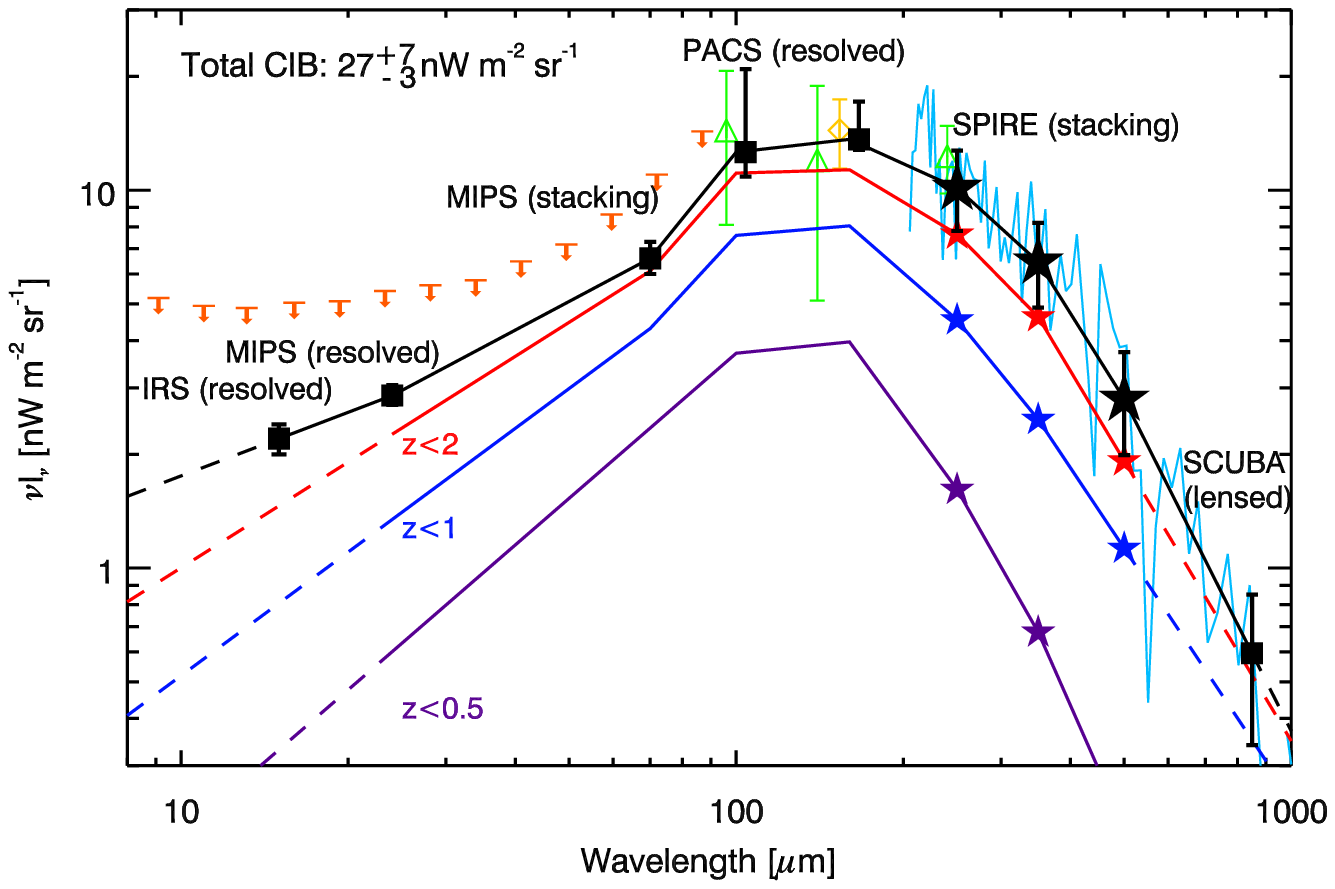}
\caption{\label{fig:CIBsynt} Spectral energy distribution of the CIB.  \textit{Black filled stars}: our total extrapolated CIB at 250\,$\micro$m, 350\,$\micro$m, and 500\,$\micro$m. \textit{Black filled squares}: total extrapolated CIB from deep number counts at 16\,$\micro$m \citep{Teplitz2011}, 24\,$\micro$m and 70$\micro$m \citep{Bethermin2010a}, 100\,$\micro$m and 160\,$\micro$m \citep{Berta2011}, and 850\,$\micro$m \citep{Zemcov2010}. \textit{Colored solid lines}: contribution of the z$<$0.5 (purple), $z<1$ (dark blue), and $z<2$ (red) sources to the CIB from the counts measured by \citet{Le_Floch2009} at 24\,$\micro$m, \citet{Berta2011} at 70\,$\micro$m, 100\,$\micro$m, and 160\,$\micro$m), and in this paper at 250\,$\micro$m, 350\,$\micro$m, and 500\,$\micro$m. \textit{Colored filled stars}: our total extrapolated CIB at 250\,$\micro$m, 350\,$\micro$m, 500\,$\micro$m for various cuts in redshift. The \textit{colored stars} indicate our new points. The \textit{dashed lines} correspond to the extrapolation of these contributions below 24\,$\micro$m and above 500\,$\micro$m. \textit{Cyan solid line:} Absolute CIB spectrum measured by \textit{COBE}/FIRAS \citep{Lagache2000}. \textit{Green triangles}: absolute CIB measurements performed by \textit{COBE}/DIRBE at 100\,$\micro$m, 140\,$\micro$m, and 240\,$\micro$m (updated in \citealt{Dole2006}). \textit{Yellow diamond:} absolute measurements of \citet{Penin2011a} at 160\,$\micro$m with \textit{Spitzer}/MIPS. \textit{Orange arrows}: upper limits derived from opacity of the Universe to TeV photons \citep{Mazin2007}. The \citet{Berta2011}, \citet{Penin2011a}, and \textit{COBE}/FIRAS points have been slightly shifted in wavelength for clarity.}
\end{figure*}

Combining the total extrapolated CIB measured from deep surveys at various wavelengths, we can produce a fully-empirical SED of the CIB (see Fig.~\ref{fig:CIBsynt}). We used the values coming from resolved counts at 16\,$\micro$m \citep{Teplitz2011}, 24\,$\micro$m \citep{Bethermin2010a}, 100\,$\micro$m and 160\,$\micro$m \citep{Berta2011}, as well as counts measured by stacking analyses at 70\,$\micro$m \citep{Bethermin2010a}, our new results at 250\,$\micro$m, 350\,$\micro$m, and 500\,$\micro$m, and also resolved sources in lensed areas at 850\,$\micro$m \citep{Zemcov2010}. From these values, we then estimate the total CIB integrated between 8\,$\micro$m and 1000\,$\micro$m: $27_{-3}^{+7}$~nW\,m$^{-2}$.sr$^{-1}$. We use power-laws to interpolate between the data points. To account for the fact that the different data points were estimated in similar fields and are thus likely to be significantly correlated, we assume a perfect correlation between each wavelengths to obtain conservative uncertainties.\\

We also derive the contribution to the total CIB from different redshift slices. We use the extrapolated values deduced from the counts per redshift slice of \citet{Le_Floch2009}, \citet{Berta2011} and our  SPIRE measurements. The \citet{Berta2011} counts were integrated following the same method as for our SPIRE counts. Fig.~\ref{fig:CIBsynt} (colored lines) shows how the CIB SED is built up as a function of redshift. The contribution of the various redshift slices to the CIB integrated between 8\,$\mu$m and 1000\,$\mu$m is given Table~\ref{tab:intCIB}.

\begin{table}
\centering
\begin{tabular}{lr}
\hline
\hline
\noalign{\vspace{1pt}}
Redshift slice & integrated CIB intensity\\
 & nW\,m$^{-2}$\,sr$^{-1}$ \\
\noalign{\vspace{1pt}}
\hline
\noalign{\vspace{1pt}}
$0<z<0.5$ & $6.3\pm1.5$\\
$0.5<z<1$ & $7.9\pm2.2$\\
$1<z<2$ & $7.7\pm2.8$\\
$z>2$ &  $4.7\pm2.0$\\
\noalign{\vspace{1pt}}
\hline
\end{tabular}
\caption{\label{tab:intCIB} Contribution of the various redshift slices to the CIB integrated between 8\,$\mu$m and 1000\,$\mu$m.}
\end{table}

\section{Discussion}

\label{sect:discussion}

\subsection{Deep source counts in the 250-500\,$\micro$m range}

With our new stacking analysis, we have confirmed, with a completely independent method, the deep counts produced by \citet{Glenn2010} using a $P(D)$ analysis. Unlike for $P(D)$ analysis, our stacking approach allows binning in redshift, providing new information on the SPIRE sources.\\

Our knowledge on the number counts in this wavelength interval has dramatically improved in few last years.  Before BLAST and \textit{Herschel}, the source counts were very poorly constrained by ground-based observations, e.g. 350\,$\micro$m had only three reported $\sim$20\,mJy sources \citep{Khan2007}.  Now, thanks to \textit{Herschel}, they are well constrained between 2\,mJy and 1\,Jy.\\

\subsection{New statistical constraints for the models}

The number counts alone are not sufficient to constrain evolution models. In this paper, we have compared different models fit to number counts. Some of these models reproduce the number counts using incorrect redshift distributions; while here we show that in fact, all the models are ruled out by our measurements. This highlights how redshift information is crucial in this context. The importance of the redshift distributions of the sources and of the CIB was also pointed out by \citet{Le_Floch2009}, \citet{Jauzac2011}, \citet{Bethermin2011} and \citet{Berta2011} among others. \citet{Le_Floch2009} measured the counts and the redshift distribution at 24\,$\micro$m with \textit{Spitzer} in the COSMOS field. \citet{Berta2011} produced a large collection of observables in the PACS bands (70\,$\micro$m, 100\,$\micro$m, and 160\,$\micro$m). Here, we provide the same type of observables in the SPIRE bands, using a stacking analysis to reach a depth similar to \citet{Berta2011}, despite having a stronger confusion. The combination of these three datasets will provide very stringent constraints for the next generation of evolution models.\\

\subsection{Origin of the sub-mm part of the cosmic infrared background}

Thanks to the depth and the precision of our new measurements, we can now study the sub-mm part of the CIB from an empirical point of view. As predicted by most of the models, the mean redshift of the CIB increases with wavelength \citep[e.g.][]{Lagache2005}. This confirms that the CIB in the sub-mm domain is dominated by the high-redshift populations. The extrapolation of our counts down to zero flux density provides an estimation of the sub-mm CIB in agreement with the absolute measurements. Our reconstruction of the properties of the SPIRE sources from the mid-infrared and optical data can thus explain how the sub-mm CIB was emitted.\\

In addition, these redshift distributions will help to interpret the CIB fluctuations measured by \textit{Herschel} (\citealt{Amblard2011}, also see calibration, flux cut and galactic cirrus discussion in \citet{Planck_CIB}) and \textit{Planck} \citep{Planck_CIB}. In fact, PACS and SPIRE redshift distributions constrain the emissivities of the infrared galaxies as a function of redshift, and will help to break degeneracies between these emissivities and the mass of the dark matter halos hosting the star-forming galaxies in the fluctuation models \citep[e.g.][]{Planck_CIB,Penin2011b}.

\section{Conclusion}

\label{sect:conclusion}

Thanks to the sensitivity of SPIRE and the high-quality of the ancillary data in the GOODS and COSMOS fields, we have determined new statistical constraints on the sub-mm galaxies. The main results of this work are:
\begin{itemize}
\item We produced deep counts (down to 2\,mJy), which confirm the previous measurements performed by stacking \citep{Bethermin2010b} and $P(D)$ analysis \citep{Glenn2010}, and significantly reduce the uncertainties in the measurements. In addition, we provide number counts per redshift slice at these wavelengths.
\item We measured the redshift distribution of the sources below the confusion limit using a stacking analysis.
\item We compared our results with the predictions of the most recent evolutionary models, which do not manage to accurately reproduce our new points. These new constraints will thus be very useful for building a new generation of models.
\item From our source counts, we also derived new estimates of the CIB level at 250\,$\micro$m, 350\,$\micro$m, and 500\,$\micro$m, in agreement and with an accuracy competitive with the FIRAS absolute measurements. We also derived constraints on the redshift distribution of the CIB.
\item Finally, combining our results with other work, we have estimated the CIB integrated between 8\,$\micro$m and 1000\,$\micro$m, produced by galaxies, to be $27_{-3}^{+7}$~nW\,m$^{-2}$\,sr$^{-1}$.
\end{itemize}

\begin{acknowledgements}
We thank the COSMOS and GOODS teams for releasing publicly their data. Thanks to Georges Helou for suggesting that the distribution of the colors is more likely log-normal than normal. MB thank Herv\'e Dole for his advices about stacking, and Elizabeth Fernandez for providing a mock catalog from the B\'ethermin et al. model. SPIRE has been developed by a consortium of institutes led by Cardiff Univ. (UK) and including Univ. Lethbridge (Canada); NAOC (China); CEA, LAM (France); IFSI, Univ. Padua (Italy); IAC (Spain); Stockholm Observatory (Sweden); Imperial College London, RAL, UCL-MSSL, UKATC, Univ. Sussex (UK); Caltech, JPL, NHSC, Univ. Colorado (USA). This development has been supported by national funding agencies: CSA (Canada); NAOC (China); CEA, CNES, CNRS (France); ASI (Italy); MCINN (Spain); SNSB (Sweden); STFC, UKSA (UK); and NASA (USA). MB acknowledge financial support from ERC-StG grant UPGAL 240039. SJO acknowledge support from the Science and Technology Facilities Council [grant number ST/F002858/1] and [grant number ST/I000976/1]. MV was supported by the Italian Space Agency (ASI ÒHerschel ScienceÓ Contract I/005/07/0). The data presented in this paper will be released through the {\em Herschel} Database in Marseille HeDaM ({hedam.oamp.fr/HerMES}).

\end{acknowledgements}

\bibliographystyle{aa}

\bibliography{biblio}

\begin{appendix}

\section{Measurement of the auto-correlation function}
\label{sect:acf_stacking}

To compute the uncertainties in our counts, we measure the auto-correlation function (ACF) using a new method, based on the stacking of a density map, containing in each pixel the number of sources centered on it. If we stack this map at the position of the sources, the expected stacked image $M(\theta)$ will be \citep{Bavouzet2008,Bethermin2010b}
\begin{equation}
M(\theta) = \left (1+w(\theta) \right ) \times \rho_{\rm S},
\end{equation}
where $\rho_{\rm S}$ is the source density (in sources\,pixel$^{-1}$) and $w(\theta)$ the ACF. Note that this method is not fully accurate because a relative error of 10$^{-3}$ on $\rho_S$ affects $w(\theta)$ by an absolute error of the same amount.\\

We can generalize this method to compute the \citet{Landy1993} estimator ($w(\theta) = \left (DD-2 \times DR +RR \right )/RR$, see Sect.~\ref{sect:ls_est}). In this case, we use two density maps, one for the real sources, and one for a simulated catalog, called hereafter "real" and "random" maps, respectively. $DD$ is estimated by stacking of the real map at the positions of the real catalog, $DR$ by stacking of the random map at the position of the real sources, and $RR$ by stacking of the random map at the position of the random sources. We then compute the \citet{Landy1993} estimator from the three stacked maps: $DD$, $DR$, and $RR$. This provides an estimate for $w(\theta,\phi)$, where $(\theta,\phi)$ are polar coordinates. To reduce the noise, we compute the mean in several annuli.\\

This method has a computation time proportional to $N_{\rm sources}$, instead of $N_{\rm sources}^2$ for the naive one. Nevertheless, the computation time is also proportional $N_{\rm pixels}^2$. A small number of pixels reduces the range of scales which can be probed. We thus use successive rebinning of our density maps to accelerate the computation of the ACF over a wide range of scales.\\

\section{Additional tables}

\begin{table*}
\centering
\begin{tabular}{lrrrr}
\hline
\hline
\noalign{\vspace{2pt}}
Flux & \multicolumn{4}{c}{Correction factor}\\
mJy & $0<z<0.5$ & $0.5<z<1$ & $1<z<2$ & $z>2$\\
\noalign{\vspace{2pt}}
\hline
\noalign{\vspace{2pt}}
\multicolumn{5}{c}{250\,$\mu$m} \\
\noalign{\vspace{2pt}}
\hline
\noalign{\vspace{2pt}}
23.8 & 0.89$\pm$0.08 & 0.79$\pm$0.08 & 0.81$\pm$0.08 & 0.86$\pm$0.09 \\
33.6 & 0.91$\pm$0.10 & 0.81$\pm$0.10 & 0.89$\pm$0.09 & 0.87$\pm$0.13 \\
47.4 & 0.92$\pm$0.15 & 0.94$\pm$0.15 & 0.82$\pm$0.14 & 0.85$\pm$0.20 \\
67.0 & 0.96$\pm$0.20 & 0.96$\pm$0.27 & 0.92$\pm$0.29 & 0.94$\pm$0.55 \\
94.6 & 1.02$\pm$0.30 & 0.96$\pm$0.65 & 1.16$\pm$0.60 & 0.89$\pm$1.04 \\
133.7 & - & 0.97$\pm$1.19 & - & - \\
188.8 & 0.96$\pm$0.83 & - & - & - \\
\noalign{\vspace{2pt}}
\hline
\noalign{\vspace{2pt}}
\multicolumn{5}{c}{350\,$\mu$m} \\
\noalign{\vspace{2pt}}
\hline
\noalign{\vspace{2pt}}
23.8 & 0.70$\pm$0.10 & 0.67$\pm$0.09 & 0.80$\pm$0.07 & 0.81$\pm$0.07 \\
33.6 & 0.82$\pm$0.15 & 0.82$\pm$0.12 & 0.73$\pm$0.10 & 0.88$\pm$0.11 \\
47.4 & 0.79$\pm$0.24 & 0.79$\pm$0.19 & 0.93$\pm$0.17 & 0.85$\pm$0.17 \\
67.0 & 0.95$\pm$0.44 & 0.67$\pm$0.37 & 0.85$\pm$0.26 & 0.86$\pm$0.35 \\
94.6 & 1.07$\pm$1.85 & 0.61$\pm$2.00 & 1.10$\pm$0.74 & - \\
133.7 & - & - & - & 0.62$\pm$1.72 \\
\noalign{\vspace{2pt}}
\hline
\noalign{\vspace{2pt}}
\multicolumn{5}{c}{500\,$\mu$m} \\
\noalign{\vspace{2pt}}
\hline
\noalign{\vspace{2pt}}
23.8 & 0.72$\pm$0.18 & 0.63$\pm$0.14 & 0.65$\pm$0.11 & 0.75$\pm$0.10 \\
33.6 & 0.74$\pm$0.32 & 0.66$\pm$0.21 & 0.77$\pm$0.18 & 0.90$\pm$0.16 \\
47.4 & - & 0.76$\pm$0.50 & 0.73$\pm$0.28 & 0.63$\pm$0.23 \\
67.0 & - & - & 0.94$\pm$1.08 & 0.85$\pm$0.86 \\
94.6 & - & - & 0.62$\pm$2.09 & 0.64$\pm$1.49 \\
\noalign{\vspace{2pt}}
\hline
\end{tabular}
\caption{\label{fig:rescf} Correction factor applied to the resolved counts in the various flux density and redshift bins.}
\end{table*}

\begin{table*}
\centering
\begin{tabular}{lrrrrrrrr}
\hline
\hline
\noalign{\vspace{2pt}}
Flux & \multicolumn{8}{c}{Completeness correction factor}\\
mJy & $0<z<0.25$ & $0.25<z<0.5$ & $0.5<z<0.75$ & $0.75<z<1$ & $1<z<1.5$ & $1.5<z<2$ & $2<z<3$ & $z>3$ \\
\noalign{\vspace{2pt}}
\hline
\noalign{\vspace{2pt}}
\multicolumn{9}{c}{250\,$\mu$m} \\
\noalign{\vspace{2pt}}
\hline
\noalign{\vspace{2pt}}
2.1 & 1.13$\pm$0.22 & 1.27$\pm$0.31 & 2.63$\pm$0.17 & 3.20$\pm$0.04 & 1.75$\pm$0.44 & 1.55$\pm$0.15 & 1.71$\pm$0.24 & 2.55$\pm$0.33 \\
3.0 & 1.05$\pm$0.13 & 1.11$\pm$0.18 & 1.81$\pm$0.28 & 2.22$\pm$0.24 & 1.35$\pm$0.32 & 1.24$\pm$0.10 & 1.36$\pm$0.19 & 1.75$\pm$0.43 \\
4.2 & 1.02$\pm$0.07 & 1.04$\pm$0.10 & 1.38$\pm$0.22 & 1.65$\pm$0.26 & 1.15$\pm$0.20 & 1.09$\pm$0.06 & 1.17$\pm$0.12 & 1.34$\pm$0.30 \\
\noalign{\vspace{2pt}}
\hline
\noalign{\vspace{2pt}}
\multicolumn{9}{c}{350\,$\mu$m} \\
\noalign{\vspace{2pt}}
\hline
\noalign{\vspace{2pt}}
2.1 & 1.26$\pm$0.22 & 1.17$\pm$0.27 & 2.11$\pm$0.35 & 1.83$\pm$0.33 & 1.61$\pm$0.37 & 1.62$\pm$0.35 & 1.72$\pm$0.25 & 9.15$\pm$6.24 \\
3.0 & 1.10$\pm$0.12 & 1.07$\pm$0.17 & 1.54$\pm$0.32 & 1.35$\pm$0.30 & 1.24$\pm$0.30 & 1.26$\pm$0.28 & 1.33$\pm$0.20 & 3.74$\pm$0.30 \\
4.2 & 1.03$\pm$0.06 & 1.02$\pm$0.09 & 1.25$\pm$0.24 & 1.14$\pm$0.21 & 1.08$\pm$0.19 & 1.10$\pm$0.17 & 1.14$\pm$0.14 & 2.06$\pm$0.32 \\
\noalign{\vspace{2pt}}
\hline
\noalign{\vspace{2pt}}
\multicolumn{9}{c}{500\,$\mu$m} \\
\noalign{\vspace{2pt}}
\hline
\noalign{\vspace{2pt}}
2.1 & 1.34$\pm$0.54 & 1.59$\pm$0.31 & 1.40$\pm$0.31 & 1.04$\pm$0.09 & 1.40$\pm$0.37 & 1.06$\pm$0.12 & 1.18$\pm$0.27 & 1.55$\pm$0.19 \\
3.0 & 1.12$\pm$0.38 & 1.25$\pm$0.22 & 1.12$\pm$0.16 & 1.08$\pm$0.04 & 1.17$\pm$0.27 & 1.02$\pm$0.06 & 1.06$\pm$0.17 & 1.24$\pm$0.13 \\
4.2 & 1.04$\pm$0.24 & 1.09$\pm$0.12 & 1.03$\pm$0.06 & 1.02$\pm$0.02 & 1.06$\pm$0.17 & 1.00$\pm$0.03 & 1.02$\pm$0.09 & 1.09$\pm$0.07 \\
\noalign{\vspace{2pt}}
\hline
\end{tabular}
\caption{\label{stackcompgoodsn}Completeness correction factor applied to the counts by stacking in GOODS-N in the various flux and redshift bins.}
\end{table*}

\begin{table*}
\centering
\begin{tabular}{lrrrrrrrr}
\hline
\hline
\noalign{\vspace{2pt}}
Flux & \multicolumn{8}{c}{Completeness correction factor}\\
mJy & $0<z<0.25$ & $0.25<z<0.5$ & $0.5<z<0.75$ & $0.75<z<1$ & $1<z<1.5$ & $1.5<z<2$ & $2<z<3$ & $z>3$ \\
\noalign{\vspace{2pt}}
\hline
\noalign{\vspace{2pt}}
\multicolumn{9}{c}{250\,$\mu$m} \\
\noalign{\vspace{2pt}}
\hline
\noalign{\vspace{2pt}}
6.0 & 1.41$\pm$0.29 & 1.20$\pm$0.25 & 1.44$\pm$0.30 & 1.49$\pm$0.28 & 1.67$\pm$0.21 & 1.24$\pm$0.36 & 1.36$\pm$0.20 & 2.41$\pm$0.07 \\
8.4 & 1.19$\pm$0.17 & 1.06$\pm$0.14 & 1.18$\pm$0.22 & 1.24$\pm$0.18 & 1.28$\pm$0.22 & 1.07$\pm$0.26 & 1.14$\pm$0.14 & 1.70$\pm$0.27 \\
11.9 & 1.09$\pm$0.09 & 1.02$\pm$0.07 & 1.06$\pm$0.13 & 1.10$\pm$0.11 & 1.12$\pm$0.15 & 1.02$\pm$0.17 & 1.05$\pm$0.08 & 1.34$\pm$0.24 \\
16.8 & 1.04$\pm$0.04 & 1.00$\pm$0.03 & 1.02$\pm$0.07 & 1.04$\pm$0.05 & 1.05$\pm$0.09 & 1.00$\pm$0.10 & 1.02$\pm$0.04 & 1.15$\pm$0.17 \\
\noalign{\vspace{2pt}}
\hline
\noalign{\vspace{2pt}}
\multicolumn{9}{c}{350\,$\mu$m} \\
\noalign{\vspace{2pt}}
\hline
\noalign{\vspace{2pt}}
6.0 & 1.40$\pm$0.37 & 1.12$\pm$0.11 & 1.17$\pm$0.15 & 1.33$\pm$0.20 & 1.52$\pm$0.27 & 1.35$\pm$0.19 & 1.67$\pm$0.28 & 3.03$\pm$0.59 \\
8.4 & 1.18$\pm$0.28 & 1.04$\pm$0.05 & 1.06$\pm$0.09 & 1.14$\pm$0.12 & 1.21$\pm$0.26 & 1.13$\pm$0.14 & 1.30$\pm$0.25 & 2.06$\pm$0.13 \\
11.9 & 1.07$\pm$0.18 & 1.02$\pm$0.02 & 1.02$\pm$0.04 & 1.05$\pm$0.06 & 1.08$\pm$0.20 & 1.05$\pm$0.08 & 1.13$\pm$0.16 & 1.59$\pm$0.35 \\
16.8 & 1.02$\pm$0.10 & 1.00$\pm$0.01 & 1.01$\pm$0.02 & 1.02$\pm$0.03 & 1.03$\pm$0.14 & 1.02$\pm$0.04 & 1.05$\pm$0.09 & 1.33$\pm$0.36 \\
\noalign{\vspace{2pt}}
\hline
\noalign{\vspace{2pt}}
\multicolumn{9}{c}{500\,$\mu$m} \\
\noalign{\vspace{2pt}}
\hline
\noalign{\vspace{2pt}}
6.0 & 1.11$\pm$0.30 & 1.05$\pm$0.18 & 1.23$\pm$0.29 & 1.20$\pm$0.35 & 1.25$\pm$0.25 & 1.14$\pm$0.25 & 1.30$\pm$0.26 & 2.22$\pm$0.52 \\
8.4 & 1.04$\pm$0.22 & 1.01$\pm$0.10 & 1.12$\pm$0.20 & 1.08$\pm$0.25 & 1.12$\pm$0.19 & 1.05$\pm$0.15 & 1.11$\pm$0.22 & 1.51$\pm$0.53 \\
11.9 & 1.02$\pm$0.15 & 1.00$\pm$0.05 & 1.06$\pm$0.14 & 1.03$\pm$0.16 & 1.06$\pm$0.14 & 1.01$\pm$0.09 & 1.04$\pm$0.16 & 1.20$\pm$0.36 \\
16.8 & 1.00$\pm$0.11 & 1.00$\pm$0.02 & 1.03$\pm$0.09 & 1.01$\pm$0.10 & 1.03$\pm$0.09 & 1.00$\pm$0.05 & 1.01$\pm$0.10 & 1.08$\pm$0.22 \\
\noalign{\vspace{2pt}}
\hline
\end{tabular}
\caption{\label{stackcompcosmos}Completeness correction factor applied to the counts by stacking in COSMOS in the various flux and redshift bins.}
\end{table*}

\begin{table*}
\centering
\begin{tabular}{lrrrrrrrr}
\hline
\hline
\noalign{\vspace{2pt}}
Flux & \multicolumn{4}{c}{$\sigma_{\rm clus}/\sigma_{\rm poi+clus}$} & \multicolumn{4}{c}{$\sigma_{\rm clus+poi}/\sigma_{\rm tot}$}\\
mJy & $0<z<0.5$ & $0.5<z<1$ & $1<z<2$ & $z>2$ & $0<z<0.5$ & $0.5<z<1$ & $1<z<2$ & $z>2$\\
\noalign{\vspace{2pt}}
\hline
\noalign{\vspace{2pt}}
\multicolumn{9}{c}{250\,$\mu$m} \\
\noalign{\vspace{2pt}}
\hline
\noalign{\vspace{2pt}}
23.8 & 0.86 & 0.85 & 0.92 & 0.77 & 0.90 & 0.90 & 0.95 & 0.86 \\
33.6 & 0.79 & 0.72 & 0.88 & 0.64 & 0.86 & 0.85 & 0.91 & 0.81 \\
47.4 & 0.65 & 0.59 & 0.68 & 0.41 & 0.81 & 0.79 & 0.83 & 0.77 \\
67.0 & 0.54 & 0.39 & 0.48 & 0.18 & 0.77 & 0.74 & 0.76 & 0.72 \\
94.6 & 0.44 & 0.17 & 0.33 & 0.11 & 0.74 & 0.72 & 0.70 & 0.72 \\
133.7 & - & 0.12 & - & - & - & 0.71 & - & - \\
188.8 & 0.17 & - & - & - & 0.72 & - & - & - \\
\noalign{\vspace{2pt}}
\hline
\noalign{\vspace{2pt}}
\multicolumn{9}{c}{350\,$\mu$m} \\
\noalign{\vspace{2pt}}
\hline
\noalign{\vspace{2pt}}
23.8 & 0.87 & 0.89 & 0.93 & 0.88 & 0.92 & 0.94 & 0.95 & 0.92 \\
33.6 & 0.80 & 0.84 & 0.84 & 0.83 & 0.88 & 0.90 & 0.91 & 0.89 \\
47.4 & 0.58 & 0.62 & 0.73 & 0.65 & 0.81 & 0.82 & 0.83 & 0.82 \\
67.0 & 0.44 & 0.26 & 0.46 & 0.36 & 0.75 & 0.78 & 0.77 & 0.76 \\
94.6 & 0.25 & 0.12 & 0.26 & - & 0.70 & 0.79 & 0.70 & - \\
133.7 & - & - & - & 0.10 & - & - & - & 0.78 \\
\noalign{\vspace{2pt}}
\hline
\noalign{\vspace{2pt}}
\multicolumn{9}{c}{500\,$\mu$m} \\
\noalign{\vspace{2pt}}
\hline
\noalign{\vspace{2pt}}
23.8 & 0.83 & 0.83 & 0.70 & 0.85 & 0.90 & 0.91 & 0.86 & 0.91 \\
33.6 & 0.60 & 0.61 & 0.52 & 0.77 & 0.82 & 0.84 & 0.80 & 0.86 \\
47.4 & - & 0.30 & 0.27 & 0.39 & - & 0.77 & 0.77 & 0.81 \\
67.0 & - & - & 0.13 & 0.19 & - & - & 0.72 & 0.74 \\
94.6 & - & - & 0.07 & 0.12 & - & - & 0.78 & 0.78 \\
\noalign{\vspace{2pt}}
\hline
\end{tabular}
\caption{\label{tab:unctab_resolved} Sources of uncertainty on the counts measured from the resolved sources. \textit{Columns 2 to 5}: relative contribution of the clustering term to the total sample variance (Poisson$+$clustering) of resolved counts ($\sigma_{\rm clus}/\sigma_{\rm poi+clus}$). \textit{Columns 6 to 9}: relative contribution the sample variance term to the total uncertainties in resolved counts ($\sigma_{\rm clus+poi}/\sigma_{\rm tot}$). $\sigma_{\rm tot}$ contains both the uncertainties in the corrections and the sample variance.}
\end{table*}

\begin{table*}
\centering
\begin{tabular}{lrrrrrrrr}
\hline
\hline
\noalign{\vspace{2pt}}
Flux & \multicolumn{4}{c}{$\sigma_{\rm clus}/\sigma_{\rm poi+clus}$} & \multicolumn{4}{c}{$\sigma_{\rm clus+poi}/\sigma_{\rm tot}$}\\
mJy & $0<z<0.5$ & $0.5<z<1$ & $1<z<2$ & $z>2$ & $0<z<0.5$ & $0.5<z<1$ & $1<z<2$ 
& $z>2$\\
\noalign{\vspace{2pt}}
\hline
\noalign{\vspace{2pt}}
\multicolumn{9}{c}{250\,$\mu$m} \\
\noalign{\vspace{2pt}}
\hline
\noalign{\vspace{2pt}}
2.1 & 0.79 & 0.95 & 0.84 & 0.77 & 0.56 & 0.63 & 0.47 & 0.44 \\
3.0 & 0.77 & 0.94 & 0.82 & 0.74 & 0.66 & 0.58 & 0.54 & 0.50 \\
4.2 & 0.75 & 0.93 & 0.80 & 0.72 & 0.74 & 0.68 & 0.64 & 0.60 \\
\noalign{\vspace{2pt}}
\hline
\noalign{\vspace{2pt}}
\multicolumn{9}{c}{350\,$\mu$m} \\
\noalign{\vspace{2pt}}
\hline
\noalign{\vspace{2pt}}
2.1 & 0.75 & 0.94 & 0.84 & 0.78 & 0.63 & 0.60 & 0.46 & 0.16 \\
3.0 & 0.73 & 0.94 & 0.81 & 0.76 & 0.66 & 0.61 & 0.60 & 0.46 \\
4.2 & 0.70 & 0.93 & 0.78 & 0.73 & 0.65 & 0.71 & 0.73 & 0.53 \\
\noalign{\vspace{2pt}}
\hline
\noalign{\vspace{2pt}}
\multicolumn{9}{c}{500\,$\mu$m} \\
\noalign{\vspace{2pt}}
\hline
\noalign{\vspace{2pt}}
2.1 & 0.69 & 0.93 & 0.77 & 0.76 & 0.44 & 0.62 & 0.55 & 0.22 \\
3.0 & 0.65 & 0.92 & 0.75 & 0.74 & 0.45 & 0.70 & 0.53 & 0.51 \\
4.2 & 0.62 & 0.91 & 0.70 & 0.71 & 0.46 & 0.74 & 0.50 & 0.51 \\
\noalign{\vspace{2pt}}
\hline
\end{tabular}
\caption{\label{tab:unctab_stacking_goodsn} \textit{Columns 2 to 5}: relative contribution of the clustering term to the total sample variance (Poisson$+$clustering) on counts measured by stacking in GOODS-N ($\sigma_{\rm clus}/\sigma_{\rm poi+clus}$). \textit{Columns 6 to 9}: relative contribution the sample variance term to the total uncertainties in resolved counts ($\sigma_{\rm clus+poi}/\sigma_{\rm tot}$). $\sigma_{\rm tot}$ contains both the uncertainties in the completeness corrections, the uncertainties in the mean colors and the scatter, and the sample variance.}
\end{table*}

\begin{table*}
\centering
\begin{tabular}{lrrrrrrrr}
\hline
\hline
\noalign{\vspace{2pt}}
Flux & \multicolumn{4}{c}{$\sigma_{\rm clus}/\sigma_{\rm poi+clus}$} & \multicolumn{4}{c}{$\sigma_{\rm clus+poi}/\sigma_{\rm tot}$}\\
mJy & $0<z<0.5$ & $0.5<z<1$ & $1<z<2$ & $z>2$ & $0<z<0.5$ & $0.5<z<1$ & $1<z<2$ 
& $z>2$\\
\noalign{\vspace{2pt}}
\hline
\noalign{\vspace{2pt}}
\multicolumn{9}{c}{250\,$\mu$m} \\
\noalign{\vspace{2pt}}
\hline
\noalign{\vspace{2pt}}
6.0 & 0.91 & 0.97 & 0.95 & 0.88 & 0.26 & 0.46 & 0.28 & 0.21 \\
8.4 & 0.89 & 0.97 & 0.93 & 0.84 & 0.39 & 0.64 & 0.40 & 0.31 \\
11.9 & 0.85 & 0.95 & 0.90 & 0.78 & 0.61 & 0.83 & 0.61 & 0.45 \\
16.8 & 0.81 & 0.93 & 0.85 & 0.69 & 0.78 & 0.80 & 0.57 & 0.42 \\
\noalign{\vspace{2pt}}
\hline
\noalign{\vspace{2pt}}
\multicolumn{9}{c}{350\,$\mu$m} \\
\noalign{\vspace{2pt}}
\hline
\noalign{\vspace{2pt}}
6.0 & 0.88 & 0.97 & 0.95 & 0.91 & 0.38 & 0.46 & 0.27 & 0.14 \\
8.4 & 0.84 & 0.96 & 0.93 & 0.87 & 0.56 & 0.69 & 0.36 & 0.19 \\
11.9 & 0.78 & 0.93 & 0.89 & 0.80 & 0.61 & 0.75 & 0.53 & 0.32 \\
16.8 & 0.68 & 0.89 & 0.84 & 0.71 & 0.39 & 0.47 & 0.37 & 0.36 \\
\noalign{\vspace{2pt}}
\hline
\noalign{\vspace{2pt}}
\multicolumn{9}{c}{500\,$\mu$m} \\
\noalign{\vspace{2pt}}
\hline
\noalign{\vspace{2pt}}
6.0 & 0.74 & 0.94 & 0.91 & 0.90 & 0.32 & 0.52 & 0.41 & 0.22 \\
8.4 & 0.60 & 0.89 & 0.87 & 0.83 & 0.25 & 0.34 & 0.41 & 0.26 \\
11.9 & 0.46 & 0.80 & 0.79 & 0.74 & 0.22 & 0.23 & 0.26 & 0.27 \\
16.8 & 0.33 & 0.64 & 0.65 & 0.61 & 0.20 & 0.16 & 0.18 & 0.21 \\
\noalign{\vspace{2pt}}
\hline
\end{tabular}
\caption{\label{tab:unctab_stacking_cosmos} Sources of uncertainty on the counts measured by stacking. \textit{Columns 2 to 5}: relative contribution of the clustering term to the total sample variance (Poisson$+$clustering) on counts measured by stacking in COSMOS ($\sigma_{\rm clus}/\sigma_{\rm poi+clus}$). \textit{Columns 6 to 9}: relative contribution the sample variance term to the total uncertainties in resolved counts ($\sigma_{\rm clus+poi}/\sigma_{\rm tot}$). $\sigma_{\rm tot}$ contains both the uncertainties in the completeness corrections, the uncertainties in the mean colors and the scatter, and the sample variance.}
\end{table*}

\end{appendix}

\end{document}